\documentclass{amsart}

\usepackage{amssymb}
\usepackage{enumerate, xspace}
\usepackage{dsfont}
\usepackage{afterpage}
\usepackage{changebar}
\usepackage[dvips]{graphicx}
\usepackage{amsthm}
\usepackage{rotating}

\numberwithin{equation}{section}

\theoremstyle{definition}

\newcommand{\be}{\begin{equation}}
\newcommand{\ee}{\end{equation}}
\newcommand{\bea}{\begin{eqnarray}}
\newcommand{\eea}{\end{eqnarray}}
\newcommand{\bs}{\begin{split}}
\newcommand{\es}{\end{split}}

\begin{document}


\title[Renormalization group method and the ionization energy theory]
  {Renormalization group method based on the ionization energy theory}
\author[Andrew Das Arulsamy]{Andrew Das Arulsamy} \email{andrew.das.arulsamy@ijs.si}

\address{Jo$\check{z}$ef Stefan Institute, Jamova cesta 39, SI-1000 Ljubljana, Slovenia}

\address{School of Physics, The
University of Sydney, Sydney, New South Wales 2006, Australia}


\begin{abstract}
Proofs are developed to explicitly show that the ionization energy theory is a renormalized theory, which mathematically exactly satisfies the renormalization group formalisms developed by Gell-Mann-Low, Shankar and Zinn-Justin. However, the cutoff parameter for the ionization energy theory relies on the energy-level spacing, instead of lattice point spacing in \textbf{k}-space. Subsequently, we apply the earlier proofs to prove that the mathematical structure of the ionization-energy dressed electron-electron screened Coulomb potential is exactly the same as the ionization-energy dressed electron-phonon interaction potential. The latter proof is proven by means of the second-order time-independent perturbation theory with the heavier effective mass condition, as required by the electron-electron screened Coulomb potential. The outcome of this proof is that we can derive the heat capacity and the Debye frequency as a function of ionization energy, which can be applied in strongly correlated matter and nanostructures. 
\end{abstract}

\maketitle


\section{Introduction}     

The ionization energy theory (IET) originated as a concept introduced in the ionization energy based Fermi-Dirac statistics ($i$FDS). The introduction of the ionization energy into the standard Fermi-Dirac statistics requires the total energy of a system to be rewritten as a function of the ionization energy~\cite{ada3,ada5}. In addition, one needs to be aware that the energy-level spacing is nothing but the difference between the first and second occupied energy levels (or the difference between the first and second ionization energies). The existence of this energy-level spacing is due to strong electron-electron repulsion \textit{and} electron-nucleus attraction. In solids, this means that the existence of band structure is due to the non-adiabatic interplay of electron-electron and electron-phonon interactions, and other subtle interactions due to electron and nuclear spins. Therefore, in the IET formalism, one can claim (for non-free-electron systems) that larger ionization energy implies stronger Coulomb repulsion between the inner (core) and outer (valence) electron, which will give rise to weakly screened (strongly interacting) electrons. As a result, this outer electron interacts strongly (via Coulomb interaction) with the inner electron to produce large energy-level spacing (in atoms) or energy gap (in solids). This means that the energy-level spacing in real solids is proportional to their constituent atomic energy-level spacing. This proportionality defines the ionization energy approximation, which has been exploited in Refs.~\cite{pop,jap,jap2,nano}. 

The claim mentioned above and its corresponding result have been proven recently~\cite{ada6}, which have given rise to the refinement of Feynman's atomic hypothesis. The original hypothesis is given in Ref.~\cite{feyn2}, while the refined statement reads, all things are made of atoms, each with unique discrete energy levels and their energy-level spacing (or ionization energies) determine how they attract or repel each other, if they are a little distance apart or squeezed close together, respectively~\cite{ada6}. The physical examples related to this refined statement are reported in Refs.~\cite{pla,cphc}.

To answer why and how such proportionality (ionization energy approximation) exists, we need to first refine another popular statement provided by Feynman: there is plenty of room at the bottom. The ``room" in this statement is not literally empty, in which it is a space with extremely strong Coulomb interaction that gives rise to specific discrete energy levels and energy-level spacing (depending on their constituent atoms). For example, if we were to place an electron in or near this space or room, all the discrete energy levels near this charge will need to be rearranged significantly and consequently, we will obtain another set of energy-level spacing due to induced polarization from the additional Coulomb interaction (again, depending on their constituent atoms). 

Secondly, the Anderson emergent phenomenon~\cite{ander2} needs to be correctly invoked within the IET. In order to do so, we need to know why does the constituent atomic energy-level spacing (a microscopic property) survived in such a way that it is proportional to the energy-level spacing of a non-free-electron solid or, why such proportionality does not exist if the solid is a free-electron metal? The latter case implies that the free-electron metallic property is in fact, an emergent property. Therefore, the ionization energy (or the proportionality of the energy-level spacing between atoms and solids) is indeed an emergent parameter for non-free-electron solids~\cite{pssb}. Whereas, the Fermi surface is an emergent parameter for free-electron solids. As such, answering these questions may lead us to have a deeper understanding of all solids, which could then direct us to the unified principles of solids.  

In the subsequent sub-sections, we will briefly introduce the classification of solids based on bonds and valence electron distribution. This is followed by some outlines and discussion to a certain extent on the general renormalization group theory (RGT)~\cite{zinn}, as well as the RGT used in high energy physics~\cite{comb}. The purpose of these discussion are to expose the difference between the standard renormalization procedure and the energy-level spacing renormalization technique from the IET. Our primary focus here is to derive and explain the energy-level spacing renormalization equations. We will also discuss the essential steps on the use of RGT in condensed matter physics based on the formalism developed by Shankar~\cite{shankar}, which is by far the most straightforward one. Other forms of RG approaches applied to condensed matter and quantum dots have been reviewed and discussed in Refs.~\cite{shankar2,sen1,shankar3,del,kaul,huh}. Subsequently, the ionization energy concept is introduced and discussed such that the readers can easily follow and understand (a) how can this concept be associated to the general RGT formalism and (b) how can it be used to study strongly correlated matter.

\subsection{Classification of solids}

Solids are composed of ion cores and electrons. The ion cores are nothing but the positively charged nuclei with strongly bound electrons that cannot be perturbed (negligible, if any) by the environment (many-body potential) in solids~\cite{ash}. On the other hand, the valence electrons are extremely sensitive to perturbations (internal and external), no matter how small. Thus, the valence electrons play the crucial role in exhibiting various exotic (superconductors~\cite{onn,bed,kami,sad,bas0,bas00,bas1,bas2,xu}, superinsulators~\cite{vino}) and other standard physical (mechanical, thermodynamical and electronic) and chemical properties of a particular solid. The structural properties (atomic arrangements) of solids are well captured by the space groups and the Bravais lattices~\cite{ash}. 

However, the crystal structure alone does not quantify the physical and chemical properties of a given solid. This inadequacy gives rise to the study of the classification of solids based on the configuration and the electronic transition probability of the valence electrons. The configuration and the distribution of the valence electrons are entirely dependent on the (i) crystal structure and (ii) the many-body wavefunction. Recall here that the transition probability for the strongly bound electrons, which we labeled collectively as the ion core earlier, is almost zero. Now, in the wavevector (\textbf{k}) space, one can classify a solid whether it is a metal or a semiconductor by studying the electron distribution, by means of the energy bands. Such classification cannot be defined clearly in real (\textbf{r}) space~\cite{ash}. 

\begin{table}

\caption{Classification of solids based on the electronic excitation probability in the absence of internal (phonon absorption) and external disturbances (temperature and/or electric field). See text for details.}
 
\begin{tabular}{l c c} 
\hline\hline 
\multicolumn{1}{l}{Group}           & Types of solids  & Electronic excitation \\  
                                    &                  & probability \\     
\hline 

\texttt{X}                          &  Free-electron & 1 \\ 
																		&  metals &	\\ \hline
                                    
\texttt{Y}                          &  Covalent and ionic & $<$ 1 \\  
																		&  bonded crystals &  \\ \hline

\texttt{Z}                          &  Molecular and hydrogen  & $<$ 1 \\  
																		 &  bonded solids &		\\ \hline

\end{tabular}
\label{Table:IA} 
\end{table}

The fundamental classifications are based on bonds (made by the valence electrons), namely, metallic, covalent, molecular and ionic bonds. There is also the so-called hydrogen-bonded solids~\cite{ash}. We can further categorize these solids: for example, we can put all the free-electron metals in group \texttt{X}, covalent and ionic crystals in group \texttt{Y}, while the molecular and hydrogen bonded solids are collected in group \texttt{Z} (see Table~\ref{Table:IA}). These characterizations are mainly due to their electronic excitation properties. This means that, one cannot excite the electrons to contribute to the electronic properties of group \texttt{Z} solids without first destroying the symmetry of their respective solid (phase transition(s) from solid to liquid and/or vapor). 

In other words, any excitation of the electrons will cause electromigration (ion diffusion) and breaking of bonds in group \texttt{Z} solids due to large ionization energy (due to localized electrons). These localized electrons contribute less to the bonding strength, hence, low melting points~\cite{LP}. Whereas, the electrons in group \texttt{Y} solids can be excited without breaking the bonds. Apparently, free-electron metals (group \texttt{X}) need not be excited (between ground and excited states) to give rise to many observed electronic and thermodynamical properties. These free electrons form a uniform Fermi gas that are completely separated from their ion cores~\cite{ash}.
It is to be noted here that the IET cannot be applied to Fermi gas (free-electron systems), whereas it can be applied to group \texttt{Y} and \texttt{Z} solids~\cite{pssb}. 

In Table~\ref{Table:IA}, free-electron metals (group \texttt{X}) do not require excitations (hence, the electronic excitation probability = 1) to conduct electricity. While the solids in the other two groups need to be excited in order to expose their physical properties (hence, the electronic excitation probability $<$ 1). In other words, group \texttt{X} metals respond to perturbations in a significantly different way than the crystals in groups \texttt{Y} and \texttt{Z}. This implies that the electrons and phonons (vibration of ion cores) in free-electron metals can be decoupled and solved separately, which will, and have given accurate physical descriptions, without losing any important information. 

The electron-phonon ($e$-$ph$) interaction that arises here always leads to $e$-$ph$ scattering (the only change is its strength). On the other hand, in the non-free-electron solids, one cannot simply decouple the electrons and the phonons, even adiabatically without losing subtle information about the $e$-$ph$ interaction. In this latter case, $e$-$ph$ interaction does not always lead to $e$-$ph$ scattering as pointed out above, but rather, there are significant changes in the effective mass, polaronic effect, spin-orbit coupling, screening and polarizability of the system. In addition, nanostructures too, especially quantum dots do not fall into group \texttt{X} due to their discrete-like energy densities (atomic-like). 

Therefore, evaluations of these physical properties (stated above) in the absence of adiabaticity between the electrons and the phonons, and in the presence of strong $e$-$e$ interaction are crucial in the field of strongly correlated matter and nanostructures. Such evaluations are also (i) often difficult, if not impossible (ii) can only be solved numerically, (iii) not straightforward due to many approximations employed simultaneously to solve them, and (iv) in many cases, the theory is not well formulated to compute anything at all~\cite{grand}. As a consequence, to tackle and overcome these issues as pointed out in (i-iv), we can employ the IET that opens up new ways of analyzing and solving the effect of strong-correlations in any doped and multi-element non-free-electron system~\cite{ada3,ada5,ada1}. 

\subsection{Renormalization group theory in high energy physics}

The specific application of the RGT in high energy physics requires removing the ultraviolet ($k \rightarrow \infty$) divergences, where $k$ is a wavenumber. Whereas, in condensed matter physics, RGT applies to the study of phase transitions of spin-dependent (magnetic) system and removing infrared ($k \rightarrow 0$) divergences. For example~\cite{comb}, in a $d$-dimensional integral,

\begin {eqnarray}
\int_0^{\Lambda}\frac{d^dk}{k^2 + m^2} = \int_0^{\Lambda}\frac{k^{d-1}dk}{k^2 + m^2} \propto \Bigg\{^{\Lambda^{d-2}~\rm{for}~d > 2}_{\ln\Lambda~\rm{for}~d = 2}, \label{eq:1000.20a}   
\end {eqnarray}

for $d < 2$ and $m = 0$, the integral behaves like

\begin {eqnarray}
\int_0^{\Lambda}\frac{k^{d-1}dk}{k^2} = \frac{k^{d-2}}{d-2}\bigg|^{\Lambda}_0, \label{eq:1000.20ab}   
\end {eqnarray}

where $\Lambda$ is a cutoff parameter in a reciprocal space ($\pi/a$). For large $\Lambda/m^2$ and $d > 2$, Eq.~(\ref{eq:1000.20a}) diverges like $\Lambda^{d-2}$ in the limit, $\Lambda \rightarrow \infty$ (ultraviolet divergence). Here, $m$ denotes mass and note that it is common in the high-energy physics literature to take $\hbar = c = 1$, where $\hbar$ and $c$ are the Planck constant divided by 2$\pi$ and the speed of light in vacuo, respectively. Therefore, $k^2 + m^2 \equiv c^2\hbar^2k^2 + m^2c^4 = E^2$. On the other hand, for $d \leq 2$ and $m = 0$, the integral diverges like $\ln\Lambda$ ($d = 2$) and $\frac{k^{d-2}}{d-2}$ ($d < 2$) as $k \rightarrow 0$ (infrared divergence). Hence, it is appropriate to start with a general formalism of RGT (given below) based on the work of Zinn-Justin~\cite{zinn} and Goldenfeld~\cite{gold}. From this general approach, we will be able to identify the relation between IET and RGT, which will be discussed in the subsequent sections. Furthermore, our statement that the IET satisfies the RGT formalism means that all the IET equations can be recast into the RG differential equation within the approach developed by Zinn-Justin, Shankar and Gell-Mann$-$Low.

\subsection{General formulation: Zinn-Justin formalism}     

The renormalization procedure originated as an empirical program to overcome the problem of infinities (singularities) in quantum electrodynamics~\cite{zinn}. The renormalization procedures have enabled one to calculate finite values from divergent expressions. In addition, the predicted finite values obtained from these renormalized results have been confirmed by experiments with high precision, which eventually gave birth to the formal RGT as it is known today~\cite{zinn}. In the subsequent sections, the RGT is introduced by means of the Hamiltonian flow. This means that the original Hamiltonian is renormalized with an arbitrary cutoff parameter, $\Lambda$, and then, in the final formula, $\Lambda$ is removed via the limits, $\Lambda \rightarrow 0$ or $\Lambda \rightarrow \infty$. All these will be clarified shortly.        

\subsubsection{Hamiltonian flow}

We first introduce the parameter, $\Lambda$, which is the dilatation parameter in such a way that the effective Hamiltonian is given by $H_{\Lambda}$. Therefore, $H \mapsto H_{\Lambda}$ where $H_{\Lambda}$ is the renormalization group (RG) transformed Hamiltonian~\cite{zinn} and $\mapsto$ denotes mapping or creating (within the Hamiltonian space) a function $H_{\Lambda}$ from $H$ that depends on $\Lambda$ explicitly. In this case, the RG idea is that the bare Hamiltonian, $H = H_{\Lambda = 1} = H_1$, while the RG transformed Hamiltonian ($H_{\Lambda}$) and for $\Lambda > 0$, we need to know the flow of the $H_{\Lambda}$. The general form of the flow equation is given by~\cite{zinn}

\begin {eqnarray}
\Lambda\frac{d}{d\Lambda}H_{\Lambda} = \mathcal{T}[H_{\Lambda}], \label{eq:1000.21a}   
\end {eqnarray}

where $\mathcal{T}$ is the transformation in the space of the Hamiltonian. We assume~\cite{zinn} that the transformation is Markovian in which, the transformation does not depend on the trajectory between $H$ and $H_{\Lambda}$. In addition, the Markovian process is stationary, meaning that the right-hand side (RHS) of Eq.~(\ref{eq:1000.21a}) is not an explicit function of $\Lambda$, but only through $H_{\Lambda}$. Moreover, the identity,

\begin {eqnarray} 
\Lambda\frac{d}{d\Lambda} = \frac{d}{d\ln\Lambda}, \nonumber
\end {eqnarray} 

will be used without warning from now onwards. Here, Eq.~(\ref{eq:1000.21a}) is a RG equation in which $\Lambda$ determines the scale changes in the dynamical process. A fixed point can be defined as independent of $\Lambda$ where we label $H^*$ as the Hamiltonian at a fixed point, in which Eq.~(\ref{eq:1000.21a}) can be written as~\cite{zinn}   

\begin {eqnarray}
\Lambda\frac{d}{d\Lambda}H^*_{\Lambda} = \mathcal{T}[H^*_{\Lambda}] = 0. \label{eq:1000.21b}   
\end {eqnarray}

Let us now work near a fixed point, in other words, we can perform linearization: $H_{\Lambda} = H^* + \Delta H_{\Lambda}$. Linearizing Eq.~(\ref{eq:1000.21a}) leads us to~\cite{zinn}

\begin {eqnarray}
\Lambda\frac{d}{d\Lambda}\Delta H_{\Lambda} = \mathcal{L}^*(\Delta H_{\Lambda}), \label{eq:1000.21c}   
\end {eqnarray}

where $\mathcal{L}^*$ is a linear operator independent of $\Lambda$. The solution to Eq.~(\ref{eq:1000.21c}) is~\cite{zinn}

\begin {eqnarray}
\Delta H_{\Lambda} = \Lambda^{\mathcal{L}^*}\Delta H_{\Lambda = 1}. \label{eq:1000.21d}   
\end {eqnarray}
 
From Eq.~(\ref{eq:1000.21d}), it is possible to study the RG flow near the fixed point where Eq.~(\ref{eq:1000.21d}) gives us the relationship between the transformed Hamiltonian, $\Delta H_{\Lambda \neq 1}$ somewhere in the neighborhood of $H^*$ and $\Delta H_{1}$, also in the neighborhood of $H^*$. The fixed points are also known as the critical (fixed) points where the fixed-point Hamiltonian ($H^*$) is invariant to $\Lambda$. For example, there are two trivial critical points for a ferromagnet, for $T = 0$ (all spins are aligned) and for $T \rightarrow \infty$ (all spins are randomly oriented). The non-trivial critical point is for $T = T_C$ where, below $T_C$, some spins are aligned that gives rise to the correlation length. The correlation length is defined as the length for a group of spins that are aligned below $T_C$. Any RG transformations must give same results at $T = 0$ (ordered) and for $T \rightarrow \infty$ (disordered). 

\textbf{ Note~1}: Every solution obtained from the IET Hamiltonian, is a renormalized solution (can be proven using the Zinn-Justin general formalism given above.)
   
\subsubsection{Classification of eigenvectors}

Assume~\cite{zinn} that $\mathcal{L}^*$ has discrete spectra with eigenvalues, $l_i$, which corresponds to a set of eigenvectors, $\mathcal{Y}_i$. Therefore, one can expand $\Delta H_{\Lambda}$ on the eigenvectors, $\mathcal{Y}_i$ of $\mathcal{L}^*$:

\begin {eqnarray}
\Delta H_{\Lambda} = \sum_ih_i(\Lambda)\mathcal{Y}_i. \label{eq:1000.21e}   
\end {eqnarray}

The flow equation, Eq.~(\ref{eq:1000.21c}) is now given by~\cite{zinn}

\begin {eqnarray}
\Lambda\frac{d}{d\Lambda}h_i(\Lambda) = l_ih_i(\Lambda), \label{eq:1000.21f}   
\end {eqnarray}

which has the solution,

\begin {eqnarray}
h_i(\Lambda) = \Lambda^{l_i}h_i(1), \label{eq:1000.21g}   
\end {eqnarray}

Here, the eigenvectors, $\mathcal{Y}_i$ can be classified according to the sign on the eigenvalues, $l_i$, provided that $\Lambda > 0$, which is true in our case because, $1 \leq \Lambda \leq \infty$. Note here that we deliberately chose the lower limit $\Lambda = 1$, and not $\Lambda = 0$ as mentioned previously. This choice does not affect the physical theory in any way since we can set any arbitrary values for $\Lambda$, which are convenient to us so as to capture the required physics accurately. Furthermore, in this case, setting $\Lambda = 0$ is meaningless as can be seen from Eq.~(\ref{eq:1000.21g}). Having explained that, now we can return to the classification issue, which are given by: 

(i) $l_i > 0$: $h_i$ and the components of $h_i$ will grow or increase with $\Lambda$, and the system moves away from the fixed point Hamiltonian, $H^*$ (repelling or unstable fixed point). In this case, $h_i$ is said to be relevant because increasing $\Lambda$ (increasing dilatation) decreases effective correlation length. For example, the Curie temperature of a ferromagnet is an unstable critical fixed point, which means that any slight deviation from this point will lead the system away, toward the stable fixed points. The stable fixed points here are for $T > T_C$ (paramagnet: disordered spins) or $T < T_C$ (ferromagnet: increasing ordering of spins).   

(ii) $l_i < 0$: $h_i$ and the components of $h_i$ decreases with $\Lambda$. Thus giving rise to irrelevant $h_i$ because larger $\Lambda$ also implies larger correlation length. Here, the system moves toward a fixed point (attractive or stable fixed point). The fixed points at $T = 0 (< T_C)$ and $T = \infty (> T_C)$ stated above are stable fixed points.     

(iii) $l_i = 0$ (vanishing eigenvalues): $h_i$ is called a marginal variable. In this case, one needs to expand Eq.~(\ref{eq:1000.21a}) beyond the linear approximation (which was carried out earlier). Hence, Eq.~(\ref{eq:1000.21f}) can be written as~\cite{zinn} 

\begin {eqnarray}
\Lambda\frac{d}{d\Lambda}h_i(\Lambda) \propto Bh_i^2(\Lambda), \label{eq:1000.21h}   
\end {eqnarray}

where $\pm B$ (constant) leads to marginally stable ($+B$) and unstable fixed points ($-B$). For $+B$ and $\Lambda \rightarrow \infty$, the solution behaves like~\cite{zinn}, 

\begin {eqnarray}
h_i(\Lambda) \propto -\frac{1}{B\ln\Lambda}. \label{eq:1000.21i}   
\end {eqnarray}

(iv) The eigenvectors that do not affect the physical properties or the components of $h_i$ are called redundant. In other words, redundant eigenvectors give only multiplicative constants.

\subsection{Coulomb screening: Shankar formalism}  

We start by first giving a simple but very important example of renormalization, where the Thomas-Fermi screened Coulomb and the bare Coulomb potentials are actually renormalized potentials~\cite{shankar}. We follow the field theory formalism provided by Shankar~\cite{shankar}, but the notations follow Ref.~\cite{ada1} for easy comparison with the IET. In the field theory approach, the Coulomb potential, after introducing a smooth cutoff, $\exp[-|\textbf{k}|/\Lambda]$ is given by~\cite{shankar}

\begin {eqnarray}
V(\textbf{k},\Lambda) = \frac{e^2}{\epsilon_0\bigg[|\textbf{k}|^2 + \frac{|\textbf{k}|}{|\textbf{k}| + \Lambda}K_s^2\bigg]}, \label{eq:1000.22}   
\end {eqnarray}
        
where $e$ denotes the electron's charge, $\epsilon_0$ is the permittivity of free space, $K_s$ is the Thomas-Fermi wavenumber and $\textbf{k}$ is the wavevector, and $|\textbf{k}| = k$ is the corresponding wavenumber. It is straightforward to note here that $V(\textbf{k},0)$ gives the Thomas-Fermi screened Coulomb potential, 

\begin {eqnarray}
V(\textbf{k},0) = \frac{e^2}{\epsilon_0(|\textbf{k}|^2 + K_s^2)}, \label{eq:1000.22a}   
\end {eqnarray}

whereas $V(\textbf{k},\infty)$ leads to the bare Coulomb potential,

\begin {eqnarray}
V(\textbf{k},\infty) = \frac{e^2}{\epsilon_0|\textbf{k}|^2}. \label{eq:1000.22b}   
\end {eqnarray}

Here, $\Lambda$ is a cutoff parameter that has a dimension of $\textbf{k}$, and it is necessary to eliminate it from the final formula, either by taking $\Lambda \rightarrow \infty$ or $\Lambda \rightarrow 0$ as carried out above. The reason for doing so is that the final formula cannot be a function of the artificially introduced cutoff parameter, $\Lambda$. In the intermediate regime, $k \ll \Lambda \ll (K_s^2/k)$, Eq.~(\ref{eq:1000.22}) is given by~\cite{shankar}

\begin {eqnarray}
V(\textbf{k},\Lambda) \approx \frac{e^2\Lambda}{\epsilon_0|\textbf{k}|K_s^2}, \label{eq:1000.23}   
\end {eqnarray}

which describes the direct proportionality between $V(\textbf{k},\Lambda)$ and $\Lambda$. This relation is very important to understand the formalism of IET with respect to RGT.

\textbf{Note~2}: We will show how the IET can be used to capture the relationship between $V(\textbf{k},\Lambda)$ and $\Lambda$. A direct proof is given that formally associates the IET with the general formulation of RGT, as discussed earlier. The renormalized screened Coulomb potential, $V_{\rm{sc}}$ plays the pivotal role with respect to strong $e$-$e$ and $e$-$ph$ interactions.   

\subsection{Renormalization group equation: Gell-Mann$-$Low formalism}     

The Gell-Mann-Low equation is known as the $\beta$-function, or the RG differential equation. This Gell-mann-Low equation is the principal result of the RG analysis~\cite{gold}. In fact, we have already introduced equations in the form of the Gell-Mann-Low equation many times earlier, which is given in Eqs.~(\ref{eq:1000.21a}),~(\ref{eq:1000.21b}),~(\ref{eq:1000.21c}),~(\ref{eq:1000.21f}) and~(\ref{eq:1000.21h}). Here, we will derive the equation in a straightforward manner. Assuming that we have a periodic lattice, an arbitrary coupling constant is expected to change or flow with respect to any change to the cutoff parameter ($\Lambda$) that controls the distance between the lattice points (lattice point spacing) or the number of lattice points considered (magnification). For example, one can write the coupling constant ($\texttt{c}$) as~\cite{alt}

\begin {eqnarray}
\texttt{c}' = \texttt{c}\Lambda^{1-\texttt{g}}, \label{eq:1000.23a}   
\end {eqnarray}

where $\texttt{g} = 1$ implies non-interacting case that leaves $\texttt{c}$ unchanged. Furthermore, for $\texttt{g} < 1$ and $\texttt{g} > 1$ one have the repulsive (unstable) and attractive (stable) fixed points, respectively. This can be understood from the classification of eigenvalues discussed above: $\texttt{g} < 1$ gives larger $\texttt{c}$ because 

\begin {eqnarray}
\texttt{c}' \propto \texttt{c}\Lambda^{1-\texttt{g}<1}. \label{eq:1000.233a}
\end {eqnarray}

On the other hand, $\texttt{g} > 1$ gives smaller $\texttt{c}$ due to 

\begin {eqnarray}
\texttt{c}' \propto \texttt{c}\Lambda^{1-\texttt{g}<0}. \label{eq:1000.2333a}
\end {eqnarray}

Usually, we need to set a cutoff value such that $\Lambda$ = $\Lambda_{\rm min} = 0$ or 1 or any finite values. Nevertheless, we will discuss this with respect to the IET in the following section. One then can iterate the RG equation, [Eq.~(\ref{eq:1000.23a})] to explore the sequence, $\texttt{c}', \texttt{c}'',~\cdot~\cdot~\cdot$, and so on. Each iteration implies a change in the distance between the lattice points (spacing) or the number of lattice points as discussed previously. For example, Eq.~(\ref{eq:1000.23a}) can be written as~\cite{alt}       

\begin {eqnarray}
&&\texttt{c}' = \texttt{c}\Lambda^{1-\texttt{g}}, \nonumber \\&&   
\ln\frac{\texttt{c}'}{\texttt{c}} = \ln\Lambda^{1-\texttt{g}}, \nonumber \\&&
\texttt{c}' = \texttt{c}e^{\ln\Lambda^{1-\texttt{g}}}. \label{eq:1000.23b}
\end {eqnarray}

Now, the flow of the coupling constant, $\texttt{c}$ with respect to $\Lambda$ is simply the differential equation given by  

\begin {eqnarray}
\int_{\texttt{c}}^{\texttt{c}'}\frac{d\texttt{c}}{\texttt{c}} = (1-\texttt{g})\int d\ln\Lambda, \label{eq:1000.23cc} 
\end {eqnarray}

\begin {eqnarray}
\frac{d\texttt{c}}{d\ln \Lambda} = \Lambda\frac{d\texttt{c}}{d\Lambda} = \texttt{c}(1-\texttt{g}) = \beta(\texttt{c}), \label{eq:1000.23c}
\end {eqnarray}

where this equation is known as the Gell-Mann-Low equation, or also called the beta ($\beta$) function. All the solutions obtained from the IET Hamiltonian through the ionization energy based Fermi-Dirac statistics satisfy equations identical to Eq.~(\ref{eq:1000.23c}). 

\section{Claim and the Proof}

From the general formulation, and the renormalized Coulomb screening potential derivation given earlier, we know that there is a relationship between the effective Hamiltonian or $V(\textbf{k},\Lambda)$ and the length scale ($\Lambda$)~\cite{zinn,gold,shankar}, though one should remove it by means of the appropriate limits [for example, see Eqs.~(\ref{eq:1000.22a}) and~(\ref{eq:1000.22b})]. In any case, one can actually expose a subtle issue here, which is: for a given system, the $V(\textbf{k},\Lambda)$ can be $\Lambda$-dependent, which makes it possible to evaluate the evolution of the Coulomb potential between two different systems. For example, substitutional doping of element C into AB (AB$_{1-x}$C$_x$) implies that $V(\textbf{k},\Lambda)$ should either be (i) dependent on $\Lambda$ explicitly (provided that the other microscopic variables such as wavefunctions and $K_s$ are fixed as constants) or (ii) the Coulomb screening is independent of $\Lambda$. If it is independent of $\Lambda$, then we need to know some details of the microscopic variables explicitly for each doping $x$.  

In IET, we have opted for the former strategy [(i) $\Lambda$-dependent], which will be proven here. However, in IET formalism, we will not employ the field-theoretic cutoff-parameter approach since $\Lambda$ is not only artificial but also microscopically not useful because it does not carry any information about the microscopic details of a particular system. For example, we will never know how to change $\Lambda$ with increasing $x$, or for different doping elements.   
\subsection {Claim}

From our earlier work, the many-electron atomic Hamiltonian is given by~\cite{ada6}

\begin {eqnarray}
\hat{H} = -\frac{\hbar^2}{2m}\sum_i \nabla_i^2 - \frac{e^2}{4\pi \epsilon_0}\sum_{i\neq j}\bigg[\frac{Z}{r_i} - \frac{1}{2} \frac{1}{|\textbf{r}_i-\textbf{r}_j|}e^{-\sigma (r_i + r_j)}\bigg], \label{eq:3.1}
\end {eqnarray} 

where $m$ is the electron's mass and $Z$ denotes the atomic number. Moreover, $|\textbf{r}_i - \textbf{r}_j| = \sqrt{r_i^2 + r_j^2 - 2r_ir_j\cos(\theta_{i,j})}$, while the labels, $i$ and $j$ identify the electrons and avoid counting them twice. The last term, which describes the screened $e$-$e$ potential makes the Hamiltonian translationally \textit{not} invariant for solids. However, the Hamiltonian can be made translationally invariant by assuming the dopants to be distributed homogeneously. For non-homogeneous distribution, we will make use of Eq.~(\ref{eq:3.3}) (Schr$\ddot{\rm o}$dinger equation with redefined eigenvalue) discussed below where we will leave Eq.~(\ref{eq:3.1}) for atomic system, as it should be. It should be noted here that the potential term in Eq.~(\ref{eq:3.1}) is the renormalized screened Coulomb potential within the IET. This means that we have renormalized the bare Coulomb potential as follows, 

\begin {eqnarray}
\frac{1}{2}\frac{1}{|\textbf{r}_1 - \textbf{r}_2|} &&\rightarrow ~~\frac{1}{2} \frac{1}{|\textbf{r}_1-\textbf{r}_2|}e^{-\sigma_1r_1}e^{-\sigma_2r_2} \nonumber\\&&  \rightarrow ~~\frac{1}{2} \frac{1}{|\textbf{r}_1-\textbf{r}_2|}e^{-\sigma(r_1 + r_2)} \Leftrightarrow \sigma_1 = \sigma_2 = \sigma, \label{eq:3.1r}
\end {eqnarray} 

where $|\textbf{r}_1 - \textbf{r}_2|$ denotes the distance between the unrenormalized (bare) electron 1 and electron 2, moreover, $e^{-\sigma_1r_1}$ represents the renormalization of electron 1 (valence) in the presence of electron 2 (core) and the nucleus. Likewise, $e^{-\sigma_2r_2}$ represents the renormalization of electron 2 (valence) in the presence of electron 1 (core) and the same nucleus. This renormalization procedure implies the condition $\sigma_1 = \sigma_2 = \sigma$ as introduced in the second line of Eq.~(\ref{eq:3.1r}). This condition also physically means that the probability is always higher to excite the valence electron first, and then followed by the excitation of the core electron. Therefore, the renormalization procedure given in Eq.~(\ref{eq:3.1r}) is different when one compares to the Thomas-Fermi (TF) screened Coulomb potential, which is given by

\begin {eqnarray}
\frac{1}{2}\frac{1}{|\textbf{r}_1 - \textbf{r}_2|} &&\rightarrow ~~\frac{1}{2} \frac{1}{|\textbf{r}_1-\textbf{r}_2|}e^{-\mu|\textbf{r}_1-\textbf{r}_2|}. \label{eq:3.1rtf}
\end {eqnarray} 

It should be clear here the reason why we did not follow the TF procedure as it considers only the distance between the screened electron 2 and electron 1, which is not in a suitable form to incorporate the energy-level spacing that originates from the electron-electron interaction due to the Coulomb attraction of both electrons toward the positive nucleus~\cite{ada6}. The $\sigma$ is given by~\cite{ada6,ada1} 

\begin {eqnarray}
\sigma = \mu e^{\frac{1}{2}\lambda(- \xi)}.\label{eq:3.2}
\end {eqnarray}
 
Here, $\mu$ is the screening constant of proportionality, while $\lambda = (12\pi\epsilon_0/e^2)a_B$, where $a_B$ denotes the Bohr radius. We will now sketch the proof required to explain how the exponential cutoff [$e^{-\sigma (r_i + r_j)}$] has come to play in Eq.~(\ref{eq:3.1}). For a detailed proof, the readers are referred to Refs.~\cite{ada6,ada1}. Its origin can be traced back to one of the restrictive conditions used in the derivation of the $i$FDS. The two standard restrictive conditions used in the Fermi-Dirac statistics are given by, (i) the total number of electrons ($N_{\rm e}$) in a given system is constant and [see Eq.~(\ref{eq:1000.24})] (ii) the total energy of all the electrons in that system is also constant [see Eq.~(\ref{eq:1000.25})]. Both conditions can be written as~\cite{ada3} 

\begin {eqnarray}
\sum_i dn_i = 0. \label{eq:1000.24}   
\end {eqnarray}

\begin {eqnarray}
\sum_i (E_{\rm electron})_i dn_i = 0. \label{eq:1000.25}   
\end {eqnarray}

Here, $dn_i$ denotes the derivative or change in the number of electrons, which must be zero because $N_{\rm e}$ is a constant. Equation~(\ref{eq:1000.24}) remains the same in both $i$FDS and the standard FDS. In Eq.~(\ref{eq:1000.25}) however, $\xi$ has been included as an additional constraint where $E_{\rm electron} = E_{\rm initial~state} + \xi$ and $E_{\rm initial~state}$ here also represents the total energy in the absence of any disturbances or when temperature is equal to zero ($E_0$). The arguments for electron-excitation is as follow: an electron to occupy a higher energy state N from the initial state M is more probable than from state L if the condition $\xi(\rm M) < \xi(\rm L)$ at certain temperature, $T$ is satisfied. Here, $E_{\rm initial~state}$ is the energy of an electron in a given system at a certain initial state that ranges from $+\infty$ to 0. For the holes, the restrictive condition reads, $E_{\rm hole} = E_{\rm initial~state} + \xi$, and the energy of a hole in a given system at a certain initial state ranges from 0 to $-\infty$. This negative range can be justified as follows: for a hole to occupy a lower state M from initial state N is more probable than to occupy state L if the condition $\xi$(M) $<$ $\xi$(L) at certain temperature, $T$ is satisfied. Therefore, Eq.~(\ref{eq:1000.25}) can be rewritten as

\begin {eqnarray}
\sum_i (E_0 \pm \xi)_i dn_i = 0. \label{eq:1000.26}
\end {eqnarray}

Using Eqs.~(\ref{eq:1000.24}) and~(\ref{eq:1000.26}) we can derive the ionization energy based Fermi-Dirac statistics for the respective electrons ($f_e$) and holes ($f_h$):~\cite{ada3}

\begin{eqnarray}
&&f_e(E_0,\xi) = \frac{1}{e^{[\left(E_0 + \xi
\right) - E_F^{(0)}]/k_BT }+1}, ~~ f_h(E_0,\xi) = \frac{1}{e^{[E_F^{(0)} - \left(E_0 - \xi
\right)]/k_BT}+1}. \label{eq:1nn}
\end{eqnarray}

Here $k_B$ is the Boltzmann constant and $E_F^{(0)}$ denotes the Fermi energy when temperature is equal to zero. Subsequently, we can calculate the changes in the electron density as a function of the energy level spacing ($\xi$) with Eq.~(\ref{eq:1000.27}) given below (we ignore the holes for convenience)

\begin{eqnarray}
&n& = \int\limits^{\infty}_0{f_e(E)D(E)dE} = 2\bigg(\frac{m}{2\lambda\pi\hbar^2}\bigg)^{3/2}
\exp\big[\lambda(E_F^{(0)} - \xi)\big]. \label{eq:1000.27}
\end{eqnarray}

Here, $D(E)$ denotes the three-dimensional density of states. The factor $e^{\lambda(E_F^{(0)} - \xi)}$ from Eq.~(\ref{eq:1000.27}) is then used to derive the static dielectric function,~\cite{ada1}  

\begin {eqnarray}
\epsilon(0,\textbf{k}) = 1 + \frac{\mathcal{K}_s^2}{|\textbf{k}|^2}\exp\big[\lambda(-\xi)\big]. \label{eq:1000.28}
\end {eqnarray}

In Eq.~(\ref{eq:1000.28}) we let $E_F^{(0)} = 0$ for convenience because it is a constant anyway (even in the presence of external or internal disturbances, for example, see Appendix: Section~\ref{A} for more details). Finally, the $\xi-$renormalized screened Coulomb potential can be derived as~\cite{ada1} 

\begin {eqnarray}
\hat{V}_{\rm{sc}} = \frac{e}{4\pi\epsilon_0r}e^{-\mu re^{\frac{1}{2}\lambda(- \xi)}} = \frac{e}{4\pi\epsilon_0r}e^{-\sigma r}. \label{eq:2.11c}
\end {eqnarray}

On the other hand, the logical proof for the existence of the relation $\sigma_1 = \sigma_2 = \sigma$ can be understood in the following way. We let $\sigma_1 = \mu_1e^{-\lambda\xi_1}$ and $\sigma_2 = \mu_2e^{-\lambda\xi_2}$. The energy level spacing ($\xi = E_{\rm electron_1} - E_{\rm electron_2}$) between electron 1 and 2 is the same for both electrons ($\xi_1 = \xi_2 = \xi$) [because we have interchanged those two electrons]. Therefore, we must satisfy these two requirements: $\mu_1 = \mu_2 = \mu$ and $\sigma_1 = \sigma_2 = \sigma$.       

In solids, we do not employ Eq.~(\ref{eq:3.1}), instead the corresponding many-body Hamiltonian (or the Schr$\rm\ddot{ o}$dinger equation) given in Eq.~(\ref{eq:3.3}) can be used~\cite{ada6}

\begin {eqnarray}
\hat{H}\varphi = (E_0 \pm \xi)\varphi. \label{eq:3.3}
\end {eqnarray}
 
The origin of $\pm\xi$ in Eq.~(\ref{eq:3.1}) is given in Eq.~(\ref{eq:1000.26}), and $\varphi$ denotes the many-body wave function. In fact, one can go on and prove that Eq.~(\ref{eq:3.3}) is equivalent to Eq.~(\ref{eq:3.1})~\cite{ada6}. Here, $\xi$ is the energy needed for a particle to overcome the bound state and the potential that surrounds it. $E_0$ denotes the total energy at temperature, $T$ = 0. We claim that every solution obtained from the IET Hamiltonian [Eq.~(\ref{eq:3.1})] is a renormalized solution, which can be recast into RG differential equation, in the form of the Gell-Mann-Low formula. In order to prove this, we first need to show that the many-electron Hamiltonian given in Eq.~(\ref{eq:3.1}) satisfies the Zinn-Justin formalism. Secondly, one also needs to prove that the Shankar Coulomb screening formula given in Eq.~(\ref{eq:1000.23}) satisfies Eq.~(\ref{eq:2.11c}). Once we have done these, then it is straightforward to follow the proof of the claim. 

\subsection {Proof}

Let us recall Eqs.~(\ref{eq:1000.21e}) and~(\ref{eq:1000.21g}), which can be rewritten as [using Eq.~(\ref{eq:3.3})]
 
\begin {eqnarray}
&&\Delta H_{\Lambda} = \sum_ih_i(\Lambda)\mathcal{Y}_i, \nonumber \\&&
h(\Lambda)\mathcal{Y} = \hat{H}\varphi. \label{eq:1010.21e}   
\end {eqnarray}

Here, $\Delta H_{\Lambda}$ is the slightly deviated Hamiltonian, from the fixed point Hamiltonian ($H^*$) via linearization procedure given earlier. In addition, $\mathcal{Y}$ and $\varphi$ are the eigenvectors, while $\Lambda$ is known as the cutoff parameter that defines the RG iterations. We are trying to rewrite the general RG Hamiltonian of Zinn-Justin with the many-electron IET Hamiltonian [Eq.~(\ref{eq:3.1})]. In this IET Hamiltonian, the screened $e$-$e$ interaction Coulomb potential is the one that has been renormalized and therefore, Eq.~(\ref{eq:1010.21e}) can be written as 

\begin {eqnarray}
&&h(\Lambda)\mathcal{Y} = \hat{V}_{\rm sc}\varphi, \nonumber \\&&
h(\Lambda) = \Lambda^{l}h(1) \Rightarrow \hat{V}_{\rm sc}(\Lambda) = \Lambda^{l}\hat{V}_{\rm unsc}, \label{eq:1011.21e}   
\end {eqnarray}
 
where $l$ is the eigenvalue ($l < 0$: because $\hat{V}_{\rm unsc}$ must decrease with $\Lambda^l$) that has its origin in the discrete linear operator, $\mathcal{L}^*$ as described in Eq.~(\ref{eq:1000.21c}). Here, $\hat{V}_{\rm unsc}$ is the unscreened (bare) Coulomb potential that gives the strongest $e$-$e$ interaction. Comparing Eq.~(\ref{eq:1011.21e}) with Eq.~(\ref{eq:2.11c}) leads to the relation

\begin {eqnarray}
\Lambda^{\rm Zinn}_{\rm Justin} = \Lambda^l = e^{-\mu re^{-\frac{1}{2}\lambda\xi}}. \label{eq:1012.21e}   
\end {eqnarray}
 
Now, taking the limit, $l \rightarrow 0$ gives $\Lambda^l \rightarrow 1$, which in turn implies $\xi \rightarrow \infty$ that leads Eq.~(\ref{eq:1011.21e}) to     
 
\begin {eqnarray}
\hat{V}_{\rm sc}(\Lambda) \rightarrow \hat{V}_{\rm unsc}, \label{eq:1014.21e}   
\end {eqnarray}
 
which is correct based on the IET. On the other hand, the limit, $l \rightarrow l_{\rm max}$ gives $\Lambda^l \rightarrow \Lambda_{\rm min}$ (because $l < 0$) that implies $\xi \rightarrow 0$ and therefore we obtain  

\begin {eqnarray}
\hat{V}_{\rm sc}(\Lambda) \rightarrow \Lambda_{\rm min}^{l_{\rm max}}\hat{V}_{\rm unsc} = \hat{V}_{\rm unsc}e^{-\mu r}, \label{eq:1015.21e}   
\end {eqnarray}

where this result is also in agreement with the IET. However, for $l \rightarrow -\infty$ gives $\Lambda^l \rightarrow 0$ that requires $\mu \rightarrow \infty$ and thus Eq.~(\ref{eq:1015.21e}) is given by
 
\begin {eqnarray}
\hat{V}_{\rm sc}(\Lambda) \rightarrow 0, \label{eq:1x15.21e}   
\end {eqnarray}
 
in which this result satisfies both IET and the original result of the Thomas-Fermi screening. 

What we have done thus far is that we have proven that the cutoff parameter, $\Lambda$ from the Zinn-Justin formalism can be directly related to the IET based screened Coulomb potential as given in Eq.~(\ref{eq:1012.21e}). In other words, $\hat{V}_{\rm sc}(\Lambda) = \hat{V}_{\rm unsc}$ is obtained for $l = 0$ ($\Lambda^l = 1$) that implies strong correlation between electrons. On the other hand, $\hat{V}_{\rm sc}(\Lambda) = \hat{V}_{\rm unsc}e^{-\mu r}$ is obtained for $l = l_{\rm max}$ ($\Lambda^l = \Lambda_{\rm min}$) and subsequently, $\hat{V}_{\rm sc}(\Lambda) = 0$ for $l \rightarrow -\infty$ ($\Lambda^l \rightarrow 0$). Therefore, the screened Coulomb potential, $\hat{V}_{\rm sc}$ is indeed a renormalized potential, without infinities. Infinities do not arise here due to the definition of Eq.~(\ref{eq:1012.21e}). Note here that the limit, $\xi \rightarrow \infty$ implies finite value for $\hat{V}_{\rm unsc}$, which in turn gives strictly finite values for both $l_{\rm max}$ and $\Lambda_{\rm min}$. For example, $\hat{V}_{\rm unsc}e^{-\mu r}$ can be related to the Thomas-Fermi screening potential by identifying $l_{\rm max} = l_{\rm TF}$ and $\Lambda^{l_{\rm TF}}_{\rm min} = \Lambda_{\rm TF} = e^{-\mu r}$. Hence, from the IET formalism, we can find the evolution of the cutoff parameter, $\Lambda$ for different doping, $x$ and doping elements. 

One can now proceed to obtain the Gell-Mann-Low equation. Using Eqs.~(\ref{eq:1011.21e}) and~(\ref{eq:1012.21e}), one can write

\begin {eqnarray}
\ln\tilde{V} = \ln\Lambda^l = -\mu re^{-\frac{1}{2}\lambda\xi}, \label{eq:1x16.21e}   
\end {eqnarray}

\begin {eqnarray}
\frac{d\ln\tilde{V}}{d\xi} = \frac{1}{2}\lambda\mu re^{-\frac{1}{2}\lambda\xi}, \label{eq:1x17.21e}   
\end {eqnarray}

\begin {eqnarray}
\frac{d\xi}{d\ln\tilde{V}} = \frac{2}{\lambda\mu re^{-\frac{1}{2}\lambda\xi}}, \label{eq:1x18.21e}   
\end {eqnarray}

where $\ln[\hat{V}_{\rm{sc}}/\hat{V}_{\rm{unsc}}] = \ln\tilde{V}$. Since $\ln\tilde{V} = \ln\Lambda^l$, we can rewrite Eq.~(\ref{eq:1x18.21e}) as  

\begin {eqnarray}
&&\frac{d\xi}{d\ln\Lambda^l} = \Lambda^l\frac{d\xi}{d\Lambda^l} = \frac{2}{\lambda\mu re^{-\frac{1}{2}\lambda\xi}} = \beta(\xi), \nonumber \\&& 
\Lambda^l\frac{d\xi}{d\Lambda^l} = \beta(\xi). \label{eq:1x19.21e} 
\end {eqnarray}

Equation~(\ref{eq:1x19.21e}) has exactly the same form as Eq.~(\ref{eq:1000.23c}). 

The next step is to prove the IET version of the screened Coulomb potential, $\hat{V}_{\rm sc}$ can be transformed into the form of the Shankar Coulomb screening formula [Eq.~(\ref{eq:1000.23})]. The Fourier transformed $\hat{V}_{\rm{sc}}$ [Eq.~(\ref{eq:2.11c})] is given by~\cite{ada1}

\begin {eqnarray}
&&\hat{V}_{\rm{sc}}(\textbf{k}) = \frac{e^2}{\epsilon_0\big(|\textbf{k}|^2 + K_s^2e^{-\lambda\xi}\big)}. \label{eq:1017.21e}
\end {eqnarray}  
 
Subsequently, one obtains    
 
\begin {eqnarray}
\hat{V}_{\rm{sc}}(\textbf{k}) = \frac{e^2}{\epsilon_0e^{-\lambda\xi}\bigg[\frac{|\textbf{k}|^2}{e^{-\lambda\xi}} + K_s^2\bigg]} \approx \frac{e^2e^{\lambda\xi}}{\epsilon_0K_s^2}. \nonumber
\end {eqnarray}   

or 
 
\begin {eqnarray}
\hat{V}_{\rm{sc}}(\textbf{k}) \approx \frac{e^2\Lambda}{\epsilon_0K_s^2}. \label{eq:1118.21e}
\end {eqnarray}

where the limit $k^2 \ll K_s^2e^{-\lambda\xi}$ has been invoked. Therefore, one obtains the relation, 

\begin {eqnarray}
\Lambda_{\rm IET} = e^{\lambda\xi}, \label{eq:1218.21e} 
\end {eqnarray}

from Eq.~(\ref{eq:1118.21e}). Comparison between Eq.~(\ref{eq:1118.21e}) and Eq.~(\ref{eq:1000.23}) gives

\begin {eqnarray}
\Lambda_{\rm Shankar} = k\Lambda_{\rm IET}. \label{eq:1019.21e}
\end {eqnarray}   
 
Hence, Eq.~(\ref{eq:1019.21e}) is the relation between the Shankar cutoff parameter and the IET-based cutoff parameter, where $\Lambda_{\rm IET}$ is a function of the ionization energy or the energy-level difference ($\xi$) of the constituent atoms in strongly correlated matter. Importantly, when $\Lambda \rightarrow \infty$, then we have $\hat{V}_{\rm{sc}}(\textbf{k}) \rightarrow \infty$, which is also in accordance with the limit, $\xi \rightarrow \infty$, as it should be. This divergence, $\hat{V}_{\rm{sc}}(\textbf{k}) \rightarrow \infty$ is entirely due to the inequalities, $k^2 \ll K_s^2e^{-\lambda\xi}$ (for $\Lambda_{\rm IET}$) and $k \ll \Lambda_{\rm Shankar} \ll (K_s^2/k)$. Having found Eq.~(\ref{eq:1019.21e}), one can go on to prove that every solution of Eq.~(\ref{eq:3.3}) is a renormalized solution. In fact, one just has to repeat the proof given below for every physical variable that has this exponential factor. However, we will only work out the proof for the $\hat{V}_{\rm{sc}}$ here, where this procedure can be easily carried out for other physical variables. 

Now one can proceed to obtain the Gell-Mann-Low equation as follows: Eq.~(\ref{eq:2.11c}) can also be written as

\begin {eqnarray}
\hat{V}_{\rm{sc}} = \frac{e}{4\pi\epsilon_0r}e^{-\mu re^{-\frac{1}{2}\lambda\xi}} = \hat{V}_{\rm unsc}e^{-\mu re^{-\frac{1}{2}\lambda\xi}} = \hat{V}_{\rm unsc}e^{-\mu r/\Lambda^{1/2}}, \label{eq:1020.21e}
\end {eqnarray}   
 
\begin {eqnarray}
\ln\frac{\hat{V}_{\rm{sc}}}{\hat{V}_{\rm{unsc}}} = -\mu re^{-\frac{1}{2}\lambda\xi} = \frac{-\mu r}{\Lambda^{1/2}}, \label{eq:1021.21e}
\end {eqnarray}   
 
where Eqs.~(\ref{eq:1118.21e}) and~(\ref{eq:1020.21e}) give us the procedure to recover the exponential function with the same mathematical structure as the Shankar smooth cutoff~\cite{shankar}, which is given by $\exp[-|k|/\Lambda]$. Note however, $\Lambda_{\rm IET}$ is always dimensionless and refers to the distance between energy levels, unlike $\Lambda_{\rm Shankar}$ that has wavenumber dimension and is based on the distance between the lattice points. Finally, one can derive the Gell-Mann-Low differential equation from Eq.~(\ref{eq:1021.21e}) as given below

\begin {eqnarray}
\frac{d\ln\tilde{V}}{d\xi} = \frac{1}{2}\lambda\mu re^{-\frac{1}{2}\lambda\xi}, \label{eq:1022.21e}
\end {eqnarray}   

\begin {eqnarray}
\frac{d\xi}{d\ln\tilde{V}} = \tilde{V}\frac{d\xi}{d\tilde{V}} = \frac{2}{\lambda\mu re^{-\frac{1}{2}\lambda\xi}} = \beta(\xi). \label{eq:1023.21e}
\end {eqnarray}   
 
Equation~(\ref{eq:1023.21e}) has exactly the same form as Eq.~(\ref{eq:1x19.21e}). Now, we can repeat the procedure between Eq.~(\ref{eq:1020.21e}) and Eq.~(\ref{eq:1023.21e}) to obtain the Gell-Mann-Low differential equation for various physical variables mentioned previously~$\blacksquare$ 

Equation~(\ref{eq:1023.21e}) also tells us that the flow of $\xi$ is determined by the changes in the physical parameter, $\tilde{V}$. From the IET, we know how the changes in the elemental composition change $\xi$ \textit{a priori}, and therefore, we can actually capture the consequential changes to the $\tilde{V}$. In other words, we do not need to know the changes occurring to $\tilde{V}$ as a result of changes in the elemental composition \textit{a priori}, in order to find $\xi$. If we can find the systematic changes to $\tilde{V}$ or any physical parameters mentioned earlier without knowing $\xi$, then we do not need IET.

Finally, from Eq.~(\ref{eq:1023.21e}), $\ln\tilde{V}$ can only be independent of $\xi$ if $\xi \rightarrow 0$ or $\xi \rightarrow \infty$. Therefore, one can obtain the corresponding $\beta$-functions that satisfy these two limits ($\xi \rightarrow 0$ and $\xi \rightarrow \infty$) as given below,

\begin {eqnarray}
\lim_{\xi \rightarrow 0} \beta(\xi_1^*) = \frac{2}{\lambda\mu r}: ~~{\rm satisfies~ Fermi~gas}, \label{eq:1024.21e}
\end {eqnarray}

\begin {eqnarray}
\lim_{\xi \rightarrow \infty} \beta(\xi_2^*) \rightarrow \infty: ~~{\rm superinsulator}. \label{eq:1025.21e}
\end {eqnarray}

As a consequence, strong electronic correlation is captured by $\beta(\xi > \xi^*_1) > 2/\lambda\mu r$.

\section{Dressed phonon and total dielectric constant}

The derivations covered in this section will be the backbone for the derivation of the acoustic and optical branches in the following section, and a further proof for the dressed phonon is given after that. Physically, $\xi$ in Eq.~(\ref{eq:3.3}) is the energy needed to excite a particular electron to a finite $r$, not necessarily $r \rightarrow \infty$. In the early stages of the ionization energy theory, we used the atomic ionization energy ($r \rightarrow \infty$) as the input parameter to compute carrier concentration~\cite{ada3}, and for this reason, $\xi$ was labeled as the ionization energy. Furthermore, Eq.~(\ref{eq:3.3}) is technically easier to use. The dressed phonon frequency is given by~\cite{ash}

\begin {eqnarray}
\omega(\textbf{k})^2 = \frac{\Omega_p^2}{\epsilon(0,\textbf{k})};~\epsilon(0,\textbf{k}) = 1 + \frac{\textsl{K}_s^2}{k^2} \label{eq:3.4}
\end {eqnarray}

where $\Omega_p$ is the plasma frequency and $\textsl{K}_s^2$ = $3n_0e^2/2\epsilon_0 E_F^0$. Here, $E_F^0$ is the Fermi level at $T$ = 0, $k$ and $\textsl{K}_s$ are the wavenumber and Thomas-Fermi wavenumber, respectively~\cite{ada1}. While, $n_0$ denotes the carrier density (free-electrons) at $T$ = 0. The electronic dielectric constant based on the Thomas-Fermi approximation and ionization energy is given by~\cite{ada1}

\begin {eqnarray}
\epsilon(0,\xi,\textbf{k}) = 1 + \frac{\textsl{K}_s^2}{k^2}\exp[\lambda(E_F^0 - \xi)]. \label{eq:3.5}
\end {eqnarray}

Using Eq.~(\ref{eq:3.5}) and taking $(\textsl{K}_s^2/k^2)\exp[\lambda(E_F^0 - \xi)] \gg 1$, we can rewrite Eq.~(\ref{eq:3.4}) as   
  
\begin {eqnarray}
\omega(\xi,\textbf{k}) = \frac{k\Omega_p}{K_s}\exp\bigg[\frac{1}{2}\lambda(\xi - E_F^0)\bigg]. \label{eq:3.6}
\end {eqnarray}

[\textbf{Note~3}: the reversible transformation between Eq.~(\ref{eq:3.4}) and~(\ref{eq:3.6}) are given in the Appendix: Section~\ref{A}]. On the other hand, the constant of proportionality ($\epsilon$) that relates the Fourier transform of the total potential ($\phi$) in a metal to the Fourier transform of the external-charge potential ($\phi_{\rm{ext}}$) can be written as~\cite{ash}

\begin {eqnarray}
\phi = \frac{1}{\epsilon}\phi_{\rm{ext}} = \frac{1}{\epsilon_{\rm{ion}}^{\rm{dressed}}}\frac{1}{\epsilon_{\rm{el}}}\phi_{\rm{ext}}, \label{eq:3.7}
\end {eqnarray}

where $\epsilon$ is defined as the total dielectric constant, $\epsilon (0,\textbf{k})$. The ionic dressed dielectric constant, $\epsilon_{\rm{ion}}^{\rm{dressed}}$ represents the screened ions due to screening electrons. We can rewrite Eq.~(\ref{eq:3.7}), after taking $(\textsl{K}_s^2/k^2)\exp[\lambda(E_F^0 - \xi)] \gg 1$ and using Eq.~(\ref{eq:3.5}) 

\begin {eqnarray}
\phi = \frac{1}{\epsilon}\phi_{\rm{ext}} = \frac{1}{\epsilon_{\rm{ion}}^{\rm{dressed}}}\frac{1}{\epsilon_{\rm{el}}}\phi_{\rm{ext}}\exp[\lambda(E_F^0 - \xi)]. \label{eq:3.7a}
\end {eqnarray}

In the presence of electronic medium with ionic medium as an external source, the potentials from the ionic ($\phi_{\rm{ion}}$) and the ionic induced electronic ($\phi_{\rm{ext}}$) contributions can be written as [using Eq.~(\ref{eq:3.5})]   

\begin {eqnarray}
\epsilon_{\rm{el}}\phi = \phi_{\rm{ion}} + \phi_{\rm{ext}}. \label{eq:3.8}
\end {eqnarray}

Similarly, considering bare ions with electronic medium as an external source, we can write~\cite{ash}

\begin {eqnarray}
\epsilon_{\rm{ion}}^{\rm{bare}}\phi = \phi_{\rm{el}} + \phi_{\rm{ext}}. \label{eq:3.9}
\end {eqnarray}

Noting that $\phi = \phi_{\rm{el}} + \phi_{\rm{ext}} + \phi_{\rm{ion}}$, and from Eqs.~(\ref{eq:3.5}),~(\ref{eq:3.7}),~(\ref{eq:3.8}) and~(\ref{eq:3.9}), we can arrive at

\begin {eqnarray}
\epsilon_{\rm{ion}}^{\rm{dressed}} = 1 + \frac{\epsilon_{\rm{ion}}^{\rm{bare}} - 1}{1 + (\textsl{K}_s^2/k^2)e^{\lambda(E_F^0 - \xi)}}. \label{eq:3.10}
\end {eqnarray}

If we take~\cite{ash} the $\epsilon_{\rm{ion}}^{\rm{bare}} = 1 - (\Omega_p^2/w^2)$, then it is straightforward to show that

\begin {eqnarray}
\frac{1}{\epsilon} = \bigg(\frac{1}{1 + (\textsl{K}_s^2/k^2)e^{\lambda(E_F^0 - \xi)}}\bigg)\bigg(\frac{\omega^2}{\omega^2 - \omega(\textbf{k})^2}\bigg). \label{eq:3.11}
\end {eqnarray}

[\textbf{Note 4}: detailed derivation of Eq.~(\ref{eq:3.11}) is given in the Appendix: Section~\ref{B}]. Equations~(\ref{eq:3.10}) and~(\ref{eq:3.11}) suggest that in order to evaluate the changes of the strongly correlated electronic properties with different doping elements, Eq.~(\ref{eq:3.5}) is sufficient. In this case, the electron-phonon interaction enters through the heavier effective mass~\cite{ada2}. However, phonons cannot be treated as a mere variation to the effective mass if we are interested in thermal properties of correlated matter.   
 
\section{Dressed acoustic and optical branches}

We consider a 1D lattice with a basis with two different ions per primitive cell and in this system of diatomic linear chain, the harmonic potential energy can be written as~\cite{ash}

\begin {eqnarray}
U_{\rm{harm}} &=& \frac{Q}{2}\sum_n[u_1(na) - u_2(na)]^2 + \frac{G}{2}\sum_n[u_2(na) - u_1((n+1)a)]^2, \label{eq:3.12}
\end {eqnarray}

where $u_1(na)$ is the displacement of the first ion that oscillates about the site $na$, while for the second ion, the displacement is $u_2(na)$ that oscillates about $na+d$. Harmonic oscillation requires $d \leq a/2$. Using Eq.~(\ref{eq:3.7a}) we can write the interaction potential constants, $Q$ and $G$ as 

\begin {eqnarray}
&&Q = \frac{\partial^2 \phi(na)}{\partial x^2}\exp[\lambda(\xi - E_F^0)]. \label{eq:3.13} \\&& 
G = \frac{\partial^2 \phi(na+d)}{\partial x^2}\exp[\lambda(\xi - E_F^0)]. \label{eq:3.14}
\end {eqnarray}

Here, $\xi$ is the average ionization energy in the vicinity of ionic mass 1 ($M_1$) and 2 ($M_2$). $\phi(na)$ and $\phi(na+d)$ represent the interaction energies between the respective ions through the screening electrons. The exponential term originates from our dressed electron described earlier. The equation of motion based on Eq.~(\ref{eq:3.12}) can be readily solved to obtain

\begin {eqnarray}
\omega^2_{\pm} &=& \frac{1}{2M_1M_2} \bigg[(Q + G)(M_1 + M_2) \pm \bigg\{\big[(Q + G)^2 \nonumber \\&& \times (M_1 + M_2)^2 - 4M_1M_2[(Q + G)^2 - Q^2 - G^2 \nonumber \\&& - 2QG\cos(ka)]\big]\bigg\}^{1/2}\bigg] e^{\lambda(\xi - E_F^0)}. \label{eq:3.15}
\end {eqnarray}

[\textbf{Note 5}: detailed derivation of Eq.~(\ref{eq:3.15}) is given in the Appendix: Section~\ref{C}]. It is clear here that the ionization-energy dress that appears in Eq.~(\ref{eq:3.15}) is identical with Eq.~(\ref{eq:3.6}). In the limit $\xi \rightarrow \infty$, then the above system is infinitely rigid. In view of Eq.~(\ref{eq:3.6}) and~(\ref{eq:3.15}), we can readily derive (following the procedure described in Ref.~\cite{ash}) the 1D harmonic oscillator Hamiltonian (in the second quantized form) and its eigenvalues, respectively as

\begin {eqnarray}
\hat{H} = \sum_\textbf{k}\hbar\omega(\textbf{k})e^{\frac{1}{2}\lambda(\xi - E_F^0)}\bigg[a_\textbf{k}^{\dagger}a_\textbf{k} + \frac{1}{2}\bigg]. \label{eq:3.15a}
\end {eqnarray}
 
\begin {eqnarray}
E = \sum_\textbf{k}\hbar\omega(\textbf{k})e^{\frac{1}{2}\lambda(\xi - E_F^0)}\bigg[n_\textbf{k} + \frac{1}{2}\bigg]. \label{eq:3.15b}
\end {eqnarray}

Note here that $n_\textbf{k}$ is the dressed phonon distribution function. The $a_\textbf{k}^{\dagger}$ and $a_\textbf{k}$ denote the usual phonon creation and annihilation operators. 

\section{Dressed electron-phonon interaction}

In the previous sections, what we have done is that we have stripped the ionization-energy dress (exponential term) from the electrons and put it on the phonons (dressed phonon). This implies that we can readily tack the exponential term to the phonon frequency and derive the Debye model and heat capacity. But before we move on, we need to prove that such discriminative dressing (stripping the electrons in order to dress the phonons) is mathematically and theoretically valid. We can prove this by showing that when we tack the exponential term (ionization-energy dress) onto the phonon frequency, we will find that the form that appears, in the derived dressed $e$-$ph$ interaction potential is exactly the same as the form that appeared in the dressed electron-electron ($e$-$e$) screened Coulomb potential (given below)~\cite{ada1}

\begin {eqnarray}
V_{\rm{ee}}(\textbf{k}) = \frac{1}{V\epsilon_0}\bigg[\frac{e^2}{\textbf{k}^2 + \textsl{K}_s^2\exp[\lambda(E_F^0 - \xi)]}\bigg]. \label{eq:3.16}
\end {eqnarray}

We know from the second-order time-independent perturbation theory~\cite{griffiths5},

\begin {eqnarray}
E^{(2)} = \sum_{n\neq m}\frac{\left|\left\langle \varphi_m^{(0)}\left|H_{\rm{per}}\right|\varphi_n^{(0)} \right\rangle\right|^2}{E^{(0)}_n - E^{(0)}_m}, \label{eq:3.17}
\end {eqnarray}
 
where $E^{(2)}$ is the second order correction to $E^{(0)}$. The $\varphi_{m,n}$ denotes wavefunction for states $m$ and $n$, respectively, while $H_{\rm{per}}$ is the perturbation. If we take $H_{\rm{per}}$ as the $e$-$ph$ interaction potential, $H_{\rm{ep}}$ for solids in \textbf{k}-space, and using Eq.~(\ref{eq:3.6}) then we can rewrite Eq.~(\ref{eq:3.17})

\begin {eqnarray}
&E^{(2)}& = \sum_{\textbf{k,k}^*}\frac{\left|\left\langle \varphi_{\textbf{k}}\left|H_{\rm{ep}}\right|\varphi_{\textbf{k}^*} \right\rangle\right|^2}{E(\textbf{k}) - E(\textbf{k}^*) - \hbar\omega(\textbf{k}-\textbf{k}^*)e^{\frac{1}{2}\lambda(\xi - E_F^0)}} \nonumber \\&& = \sum_{\textbf{k,k}^*}n_\textbf{k}(1 - n_{\textbf{k}^*}) \bigg[\frac{\left|\left\langle \varphi_{\textbf{k}}\left|H_{\rm{ep}}\right|\varphi_{\textbf{k}^*} \right\rangle\right|^2}{E(\textbf{k}) - E(\textbf{k}^*) - \hbar\omega(\textbf{k}-\textbf{k}^*)e^{\frac{1}{2}\lambda(\xi - E_F^0)}}\bigg]. \label{eq:3.18}
\end {eqnarray}

The detailed reasons why Eq.~(\ref{eq:3.17}) represents the $e$-$ph$ interaction and the missing first-order correction (its matrix elements equal zero due to orthogonality) are given in Ref.~\cite{grim}. The dressed phonon distribution function can be written as 

\begin {eqnarray}
n_\textbf{k} = \frac{1}{e^{\beta \hbar\omega(\textbf{k})e^{\frac{1}{2}\lambda(\xi - E_F^0)}} - 1}. \label{eq:3.19}
\end {eqnarray}

The denominator in Eq.~(\ref{eq:3.18}) have made use of the crystal momentum conservation, $\textbf{k} - \textbf{k}^* = \textbf{q}$, and $\beta = 1/k_BT$. Using Landau approach~\cite{ash}, we can derive the $e$-$ph$ interaction potential, $V_{\textbf{k},\textbf{k}^*}$ from Eq.~(\ref{eq:3.18}) 

\begin {eqnarray}
&&V_{\textbf{k},\textbf{k}^*} = \frac{\partial^2E^{(2)}}{\partial n_\textbf{k}\partial n_{\textbf{k}^*}} = \left|g_{\textbf{k},\textbf{k}^*}\right|^2\bigg[\frac{2\hbar\omega(\textbf{k}-\textbf{k}^*)e^{\frac{1}{2}\lambda(\xi - E_F^0)}}{[\hbar\omega(\textbf{k}-\textbf{k}^*)e^{\frac{1}{2}\lambda(\xi - E_F^0)}]^2 - [E(\textbf{k}) - E(\textbf{k}^*)]^2}\bigg]. \label{eq:3.20}
\end {eqnarray}

The $e$-$ph$ coupling constant, $ \left|g_{\textbf{k},\textbf{k}^*}\right|^2 = \left|\left\langle \varphi_{\textbf{k}}\left|H_{\rm{ep}}\right|\varphi_{\textbf{k}^*} \right\rangle\right|^2$ is given by~\cite{ash}

\begin {eqnarray}
\left|g_{\textbf{k},\textbf{k}^*}\right|^2 = \frac{1}{V}\frac{e^2}{\epsilon_0(|\textbf{k}-\textbf{k}^*|^2 + K_s^2)}\frac{1}{2}\hbar\omega(\textbf{k}-\textbf{k}^*). \label{eq:3.21}
\end {eqnarray}

In the previous work~\cite{ada2}, we have claimed that the ionization energy dressed electron-electron screened Coulomb potential has the electron-phonon interaction effect taken into account implicitly that gives rise to heavier effective mass. Consequently, we need to impose this condition of heavier effective mass into Eq.~(\ref{eq:3.20}). Here, we will make use of the effective mass ($m^*$) theorem~\cite{ash}, in which it is approximately given by 

\begin {eqnarray}
\frac{\partial^2E_n(\textbf{k})}{\partial \textbf{k}_i\partial \textbf{k}_j} \propto \frac{1}{m^*}\delta_{i,j} + \textsl{O}\bigg(\frac{1}{m^*}\bigg)^2. \label{eq:3.22} 
\end {eqnarray}

This means that for heavier effective mass, we need to satisfy, $[\hbar\omega(\textbf{k}-\textbf{k}^*)e^{\frac{1}{2}\lambda(\xi - E_F^0)}]^2 \gg [E(\textbf{k}) - E(\textbf{k}^*)]^2$. In other words, we need the term, $[E(\textbf{k}) - E(\textbf{k}^*)]^2$ to be relatively small in order to comply with Eq.~(\ref{eq:3.22}), which gives rise to the heavier effective mass effect. As such, Eq.~(\ref{eq:3.20}) can now be rewritten as (also, after dressing Eq.~(\ref{eq:3.21}) with the ionization energy) 

\begin {eqnarray}
V_{\rm{ep}}(\textbf{k},\textbf{k}^*) = \frac{1}{V\epsilon_0}\bigg[\frac{e^2}{|\textbf{k}-\textbf{k}^*|^2 + K_s^2\exp[\lambda(E_F^0 - \xi)]}\bigg].\label{eq:3.23}
\end {eqnarray}

As anticipated, Eq.~(\ref{eq:3.23}) is in exact form with Eq.~(\ref{eq:3.16}). This completes the proof that the ionization energy is proportional to the electron-phonon coupling constant and its relationship with heavier effective mass, where we have only explained this proportionality qualitatively in Ref.~\cite{ada2}. [\textbf{Note 6}: detailed derivation of Eq.~(\ref{eq:3.23}) is given in the Appendix: Section~\ref{D}]. The next step is to apply the outcome of this proof to derive the heat capacity and the well known Debye model (Debye frequency and temperature). 

\section{Dressed heat capacity and Debye temperature}

Now, we will make use of all the results derived earlier to derive the dressed heat capacity formula, dressed Debye frequency and the dressed Debye temperature. The energy density of a harmonic crystal is given by (using Eq.~(\ref{eq:3.15b}) and introducing the static potential energy constant) 

\begin {eqnarray}
&U& = U_{\rm{stat}} + \frac{1}{V}\sum_{\textbf{k},s}\frac{1}{2}\hbar\omega_s(\textbf{k})e^{\frac{1}{2}\lambda(\xi - E_F^0)} + \frac{1}{V}\sum_{\textbf{k},s}\frac{\hbar\omega_s(\textbf{k})e^{\frac{1}{2}\lambda(\xi - E_F^0)}}{e^{\beta\hbar\omega_s(\textbf{k})e^{\frac{1}{2}\lambda(\xi - E_F^0)}} - 1},\label{eq:3.24}
\end {eqnarray}

and the non-constant heat capacity ($C_v$) is given by 

\begin {eqnarray}
C_v = \frac{\partial U}{\partial T} = \frac{1}{V}\sum_{\textbf{k},s}\frac{\partial}{\partial T}\frac{\hbar\omega_s(\textbf{k})e^{\frac{1}{2}\lambda(\xi - E_F^0)}}{e^{\beta\hbar\omega_s(\textbf{k})e^{\frac{1}{2}\lambda(\xi - E_F^0)}} - 1}. \label{eq:3.250}
\end {eqnarray}

We also have introduced the branch index, $s$ = 1,...,3$p$, where $p$ is the number of ions in the basis signifying that for each $\textbf{k}$ there are 3$p$ normal modes~\cite{ash}. In the limit of long wavelength (small $k$), $\omega(\textbf{k}) = c_s(\hat{\textbf{k}})ke^{\frac{1}{2}\lambda(\xi - E_F^0)}$ [using Eq.~(\ref{eq:3.6})] where $c_s$ is the long wavelength phase velocity that depends on the vibrational polarization, $s$ and the direction of \textbf{k}~\cite{ash}. Again, using the same procedure given in Ref.~\cite{ash}, we solve Eq.~(\ref{eq:3.250}) and obtain

\begin {eqnarray}
C_v = \frac{2\pi^2k_B}{5} \bigg[\frac{k_BT}{\hbar c}\bigg]^3 e^{-\frac{3}{2}\lambda(\xi - E_F^0)}. \label{eq:3.26}
\end {eqnarray}
 
Note here that Eq.~(\ref{eq:3.26}) correctly gives the inverse proportionality between $C_v$ and the ionization energy, $\xi$. Later, this proportionality will be associated to the stiffness of the crystal, as we can anticipate from this relationship itself. [\textbf{Note 7}: detailed derivation of Eq.~(\ref{eq:3.26}) is given in the Appendix: Section~\ref{E}].

Next, let us work within Debye's approach and derive the Debye frequency, temperature and the heat capacity, all dressed with ionization energy. In this case, we may simply write, 

\begin {eqnarray}
&&\omega = cke^{\frac{1}{2}\lambda(\xi - E_F^0)},~~ k = \frac{\omega}{c}e^{-\frac{1}{2}\lambda(\xi - E_F^0)}, \label{eq:3.27}
\end {eqnarray}

where $c$ is the constant sound velocity. The phonon density of states, $N(\omega)$ is given by

\begin {eqnarray}
&N(\omega)& = \frac{dN}{d\omega} = \frac{d}{d\omega}\bigg[\frac{V\omega^3}{6\pi^2c^3}e^{-\frac{3}{2}\lambda(\xi - E_F^0)}\bigg] = \frac{\omega^2V}{2\pi^2c^3}e^{-\frac{3}{2}\lambda(\xi - E_F^0)}, \label{eq:3.28}
\end {eqnarray}
      
where $N$ is the number of phonons. By defining $\omega$ = $\omega_{\rm{D}}$ and $n_{\rm{ph}} = N/V$ as phonon density, we can rewrite Eq.~(\ref{eq:3.28}) to obtain the Debye frequency

\begin {eqnarray}
\omega_{\rm{D}} = c[6\pi^2n_{\rm{ph}}]^{1/3}e^{\frac{1}{2}\lambda(\xi - E_F^0)}. \label{eq:3.29}
\end {eqnarray}
 
Also noting that $N/V = [k_{\rm{D}}e^{-\frac{1}{2}\lambda(\xi - E_F^0)}]^3/6\pi^2$ and $k_B\Theta_{\rm{D}} = \hbar\omega_{\rm{D}} = \hbar c k_{\rm{D}}$, then we can arrive at the Debye temperature 

\begin {eqnarray}
\Theta_{\rm{D}} = \frac{\hbar c}{k_B}[6\pi^2n_{\rm{ph}}]^{1/3}e^{\frac{1}{2}\lambda(\xi - E_F^0)}. \label{eq:3.30}
\end {eqnarray}

[\textbf{Note 8}: detailed derivation of Eq.~(\ref{eq:3.30}) is given in the Appendix: Section~\ref{F}]. Before we continue, it is important to reduce the level of confusion that may arise in the above formulations and proofs with the exponential term appearing here and there. Hence, we will give a step-by-step derivation for the Debye heat capacity, which will prove the origin of the exponential term in Eq.~(\ref{eq:3.30}), as well as point out the correct way of handling this exponential term. Again, we start from Eq.~(\ref{eq:3.250}) using~\cite{ash} $d\textbf{k} = k^2dkd\Omega$ to get ($\Omega$ is the phonon polarization volume in \textbf{k}-space) 

\begin {eqnarray}
C_v = \frac{3(\hbar c)^2}{2\pi^2k_BT^2}\int^{k_{\rm{D}}}_0 k^4dke^{\lambda(\xi - E_F^0)} \frac{e^{\beta\hbar cke^{\frac{1}{2}\lambda(\xi - E_F^0)}}}{[e^{\beta\hbar cke^{\frac{1}{2}\lambda(\xi - E_F^0)}} - 1]^2}. \label{eq:3.31}
\end {eqnarray}

Now we make the substitution, $x = \beta\hbar cke^{\frac{1}{2}\lambda(\xi - E_F^0)}$ into Eq.~(\ref{eq:3.31}) and obtain

\begin {eqnarray}
C_v = 9 k_B n_{\rm{ph}}\bigg[\frac{T}{\Theta_{\rm{D}}e^{\frac{1}{2}\lambda(\xi - E_F^0)}}\bigg]^3 \int^{x_{\rm{D}}}_0\frac{x^4e^x}{(e^x - 1)^2}dx. \label{eq:3.32}
\end {eqnarray}
 
We can see that Eq.~(\ref{eq:3.32}) is in agreement with Eq.~(\ref{eq:3.26}) and~(\ref{eq:3.30}). [\textbf{Note 9}: detailed derivation of Eq.~(\ref{eq:3.32}) is given in the Appendix: Section~\ref{G}]  

\section{Remarks on quantum phase transitions and phase diagram}

\begin{figure}[hbtp!]
\begin{center}
\scalebox{0.6}{\includegraphics{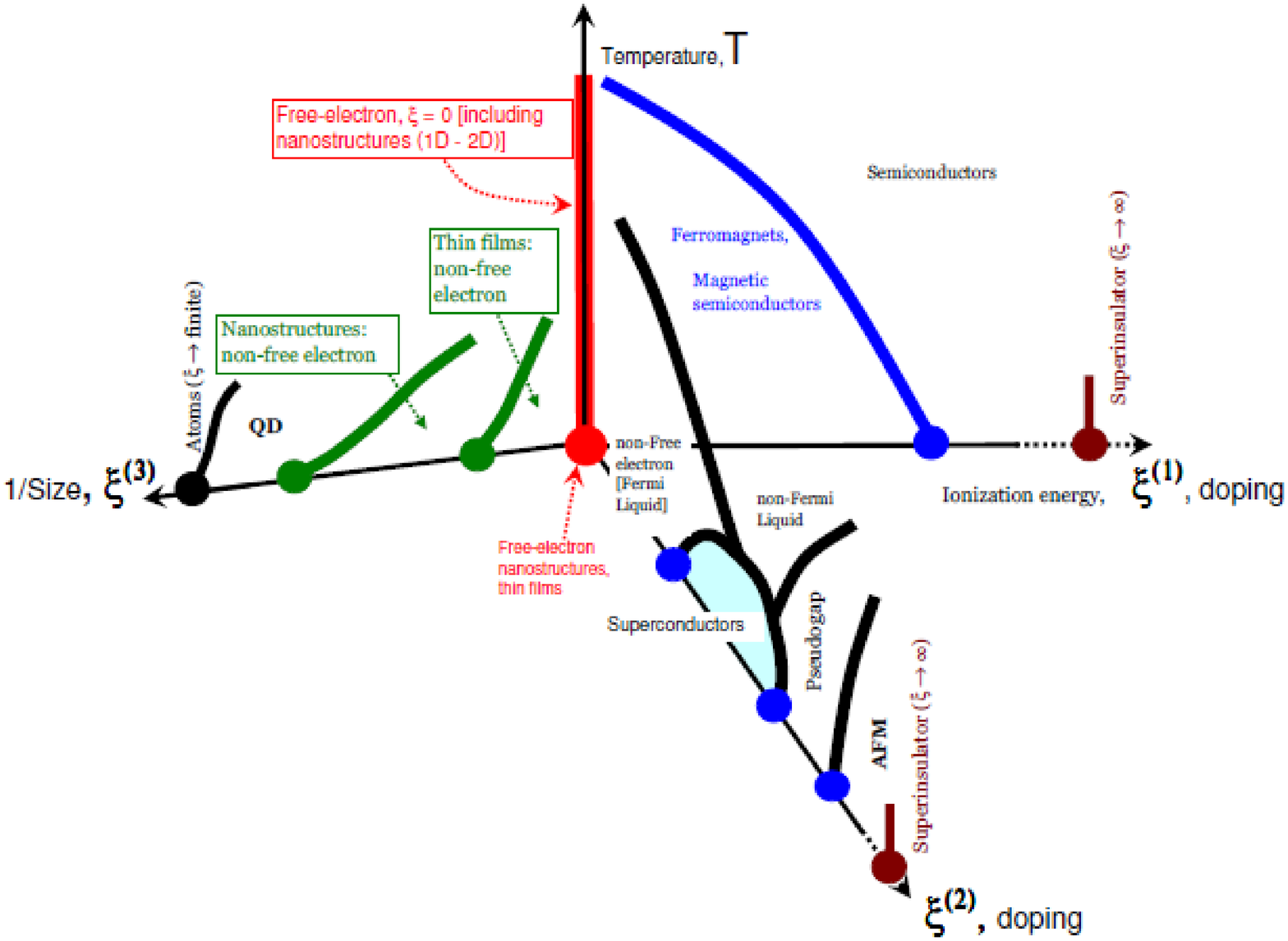}}
\caption{\large Guessed multi-dimensional phase diagram based on the ionization energy theory (IET). All the regions moving radially outward from the temperature axis (free electron metals, $\xi = 0$) are labeled as strongly correlated systems (non-free electron systems). IET cannot be applied to the free-electron metals. The critical points (filled circles) define the possible phase transitions within the strongly correlated system. The changes in the ionization energy, $\xi^{(1)}$ and $\xi^{(2)}$ are due to doping for ferromagnets and superconductors, respectively. Whereas, the left-hand side axis is due to decrease in the size and the system dimensionality: from 3-dimensional thin films (or bulk) to one- and two-dimensional nanostructures to zero-dimensional QDs. The limits, $\xi \rightarrow 0$ (free-electron metal) and $\xi \rightarrow \infty$ (superinsulator) denote two critical points from IET. Any phase transition that may occur between $\xi = 0$ and $\infty$ can be either first-order or continuous phase transition. Along the $\xi^{(2)}$ axis, the label AFM denotes antiferromagnets.}   
\label{fig:XYZ}
\end{center}
\end{figure}

Here, we briefly discuss the association between the IET and the quantum phase transitions with a multi-dimensional schematic phase diagram given in Fig.~\ref{fig:XYZ}. The graphical structure of various phase transitions given in Fig.~\ref{fig:XYZ} can be understood from the discussion that follows. The growth of thin films starting from a single atom, which cluster together to form clusters or quantum dots (QDs), eventually ($N_{\rm{atom}} \rightarrow \infty$) will either form (i) a free-electron thin film if the atoms are, for example, Cu, or (ii) a semiconducting thin film, if the atoms are Si. If there are more than one atomic species (Cu, Mn, O, etc.), say, manganites or cuprates, then these thin films or crystals can be categorized as a strongly-correlated system, not solely because they exhibit phase transitions (ferromagnetism or superconductivity), but because they no longer satisfy free-electron metallic system. Apparently, there is a continuous phase transition, from an atomic system to either free-electron metals or strongly correlated matter, if one takes the thermodynamical limit, $N_{\rm{atom}} \rightarrow \infty$, and depending whether $\xi = 0$ or $>$ 0.     

The ionization energy-based exponential function describes the quantum phase transition due to the definition of the ionization energy, which is directly related to the excitation probability function. This exponential function is also a universal function since it is independent of most of the microscopic many-body details, for example, the Hamiltonian is strictly dependent on the screened Coulomb potential while, the many-body potential is taken to be a constant. Hence, the Shankar cutoff parameter~\cite{shankar}, $\Lambda_{\rm Shankar}$ ($\textbf{k}-$dependent: distance between lattice points in the reciprocal space) has been recaptured via $\Lambda_{\rm IET}$ ($\xi-$dependent: distance between energy levels). Further investigations with respect to Sachdev model~\cite{sachdev,sachdev2,sachdev3,sachdev4,sachdev5} are anticipated to be of particular interest. 

Apart from that, it is interesting to note that the continuous transition from Mott insulator to Fermi liquid developed by Senthil~\cite{senthil1,senthil2} for 2D organic material is somewhat qualitatively related to our discussion of the phase transition from free-electron metals to strongly correlated systems. This implies the possibility to develop a theory of strong correlation above the Mott-Hubbard transition temperature. 

Moreover, we also anticipate that chemical reactions (for example, from two unreacted chemical species) must involve energy-level crossings and are expected to undergo continuous quantum phase transition~\cite{scr}. If the formed compound after chemical reactions, turned out to be a non-Fermi-electron type, then one can indeed apply the IET to predict numerous physical, thermodynamical and chemical properties.  

\section{Conclusions}

In conclusion, we have proved that the ionization energy theory is indeed a renormalized theory based on the approach developed by Zinn-Justin, Shankar and Gell-Mann$-$Low formalism. Subsequently, we have applied these proofs to show that the ionization-energy dressed electron-phonon interaction potential is exactly in the same form as the ionization-energy dressed electron-electron Coulomb potential. Consequently, we can obtain the heat capacity and the Debye frequency as a function of ionization energy. These latter proofs can be used to explain the evolution of the electron-phonon interaction with respect to elemental composition in strongly correlated matter. Apart from that, this theory may define the existence of the (quantum) phase transition between free-electron metals and strongly correlated matter, which will be addressed in our future work.

\section*{Acknowledgments}

This work was supported by the Slovene Human Resources Development and Scholarship Fund (Ad-Futura), the Slovenian Research Agency (ARRS) and the Institut Jo$\check{z}$ef Stefan (IJS). I also would like to thank the School of Physics, University of Sydney for the USIRS award (2007$-$2009). I am extremely grateful to one of the anonymous referees for the excellent comments and also for correcting all of my mistakes. I gratefully acknowledge the financial support provided by the late Madam Kithriammal Soosay who passed away suddenly while this article was under review. Her unconditional and continuous support during the period 1999$-$2010 was paramount to the development of the ionization energy theory.   

\section{Appendix}

\subsection{Reversible transformation between Eq.~(\ref{eq:3.4}) and~(\ref{eq:3.6})}\label{A}

One should not assume that Eq.~(\ref{eq:3.4}) and Eq.~(\ref{eq:3.6}) are theoretically different. Equations~(\ref{eq:3.4}) and Eq.~(\ref{eq:3.6}) are both mathematically and theoretically exact, which can be transformed and reverse-transformed exactly. We have to prove this transformation because obtaining Eq.~(\ref{eq:3.6}) from Eq.~(\ref{eq:3.4}) is our starting axiom for the later proofs in this appendix. Hence, by taking the appropriate limits,  

\begin {eqnarray}
\lim_{\xi > E_F^0}\epsilon(0,\xi,\textbf{k}) = 1 + \frac{\textsl{K}_s^2}{k^2}\exp[\lambda(E_F^0 - \xi)], \nonumber
\end {eqnarray}

\begin {eqnarray}
\lim_{\xi \rightarrow E_F^0}\epsilon(0,\xi,\textbf{k}) = 1 + \frac{\textsl{K}_s^2}{k^2} = \epsilon(0,\textbf{k}), \nonumber
\end {eqnarray}

and also after taking, $(\textsl{K}_s^2/k^2)\exp[\lambda(E_F^0 - \xi)] \gg 1$ and $\textsl{K}_s^2/k^2 \gg 1$, we obtain

\begin {eqnarray}
\lim_{\xi > E_F^0}\omega(\xi,\textbf{k}) = \frac{k\Omega_p}{K_s}\exp\bigg[\frac{1}{2}\lambda(\xi - E_F^0)\bigg], \nonumber
\end {eqnarray}

\begin {eqnarray}
\lim_{\xi \rightarrow E_F^0}\omega(\xi,\textbf{k}) = \frac{k\Omega_p}{K_s} = \omega(\textbf{k}) = \frac{\Omega_p}{\sqrt{\epsilon(0,\textbf{k})}}. \nonumber
\end {eqnarray}

For conventional free-electron metals, the $\lim_{\xi \rightarrow E_F^0}$ applies and any perturbation due to temperature and/or potential (due to doping, defects, etc.) will vary the $E_F^0$ and $n_0$ accordingly. For example, for each type of free-electron metal, and for each type of perturbation, there are $n_0$ and $E_F^0$, which are both unique. In our approach, any variations to $E_F^0$ and $n_0$ are captured through $\xi$. Meaning, for a given system, $E_F^0$ and $n_0$ are constants, while $\xi$ will respond to any perturbations, be it large or small. For example, $n(\textbf{r},T,\rm{doping})$ = $n_0 \exp[\lambda(E_F^0 - \xi)]$ that will tell us the changes to $n(\textbf{r},T,\rm{doping})$. In other words, $n(\textbf{r},T,\rm{doping})$ = $n_0(\rm{of~that~perturbed~system})$, or simply, $\lim_{\xi \rightarrow E_F^0} n(\textbf{r},T,\rm{doping})$ = $n_0(\rm{of~that~perturbed~system})$, which will take us back to square one, the free-electron metals~\cite{ada1}. As a consequence, we can see that $\omega(\xi,\textbf{k})$ = $\omega(\textbf{k})$ or Eq.~(\ref{eq:3.4}) and Eq.~(\ref{eq:3.6}) are exactly the same due to our reversible transformation. Simply put, the $\lim_{\xi > E_F^0}$ implies $E_F^0$ and $n_0$ have been taken as constants for a given system and we are letting $\xi$ to respond accordingly to any external and/or internal perturbations. 

\subsection{Derivation of Eq.~(\ref{eq:3.11})}\label{B}

We start from the definition as given in Eq.~(\ref{eq:3.7}) where,

\begin{eqnarray}
\epsilon\phi = \phi_{\rm{ext}}. \label{eq:3.1AA} 
\end{eqnarray}

Next, we add Eq.~(\ref{eq:3.8}) and Eq.~(\ref{eq:3.9}) to obtain,

\begin{eqnarray}
\phi(\epsilon_{\rm{el}} + \epsilon_{\rm{ion}}^{\rm{bare}}) = 2\phi_{\rm{ext}} + \phi_{\rm{ion}} + \phi_{\rm{el}}. \label{eq:3.2AA} 
\end{eqnarray}

We then subtract Eq.~(\ref{eq:3.1AA}) from Eq.~(\ref{eq:3.2AA}) to get

\begin{eqnarray}
\phi(\epsilon_{\rm{el}} + \epsilon_{\rm{ion}}^{\rm{bare}} - \epsilon) = \phi_{\rm{ext}} + \phi_{\rm{ion}} + \phi_{\rm{el}}. \label{eq:3.3AA} 
\end{eqnarray}

Since, $\phi = \phi_{\rm{ext}} + \phi_{\rm{ion}} + \phi_{\rm{el}}$, Eq.~(\ref{eq:3.3AA}) needs to satisfy

\begin{eqnarray}
\epsilon = \epsilon_{\rm{el}} + \epsilon_{\rm{ion}}^{\rm{bare}} - 1. \label{eq:3.4AA} 
\end{eqnarray}

Now, if the electrons screen the dressed ions, then Eq.~(\ref{eq:3.1AA}) can also be written as

\begin{eqnarray}
(\epsilon_{\rm{ion}}^{\rm{dressed}}\epsilon_{\rm{el}})\phi = \phi_{\rm{ext}}, \label{eq:3.5AA} 
\end{eqnarray}  

such that [from Eq.~(\ref{eq:3.5AA}) and Eq.~(\ref{eq:3.4AA})]
  
\begin{eqnarray}
\epsilon_{\rm{ion}}^{\rm{dressed}}\epsilon_{\rm{el}} = \epsilon_{\rm{el}} + \epsilon_{\rm{ion}}^{\rm{bare}} - 1 = \epsilon. \label{eq:3.6AA} 
\end{eqnarray}  

Therefore, 

\begin{eqnarray}
\epsilon_{\rm{ion}}^{\rm{dressed}} = 1 + \frac{\epsilon_{\rm{ion}}^{\rm{bare}} - 1}{\epsilon_{\rm{el}}}. \label{eq:3.7AA} 
\end{eqnarray}  
 
Substituting Eq.~(\ref{eq:3.5}) into Eq.~(\ref{eq:3.7AA}) will lead us directly to Eq.~(\ref{eq:3.10}). Subsequently, we substitute $\epsilon_{\rm{ion}}^{\rm{bare}} = 1 - (\Omega_p^2/\omega^2)$ and Eq.~(\ref{eq:3.6}) into Eq.~(\ref{eq:3.7AA}) so as to arrive at 

\begin{eqnarray}
&\epsilon_{\rm{ion}}^{\rm{dressed}} &= 1 + \frac{\epsilon_{\rm{ion}}^{\rm{bare}} - 1}{\epsilon_{\rm{el}}} \nonumber \\&& = 1 + \frac{1 - (\Omega_p^2/\omega^2) - 1}{\epsilon_{\rm{el}}} \nonumber \\&& = 1 - \frac{(\Omega_p^2/\omega^2)}{\epsilon_{\rm{el}}} \nonumber \\&& = 1 - \frac{(\Omega_p^2/\epsilon_{\rm{el}})}{\omega^2} \nonumber \\&& = 1 - \frac{\omega(\textbf{k})^2}{\omega^2}. \label{eq:3.8AA} 
\end{eqnarray} 

From Eq.~(\ref{eq:3.6AA}), we can rewrite the total dielectric, $\epsilon$ as

\begin{eqnarray}
\frac{1}{\epsilon} = \frac{1}{\epsilon_{\rm{ion}}^{\rm{dressed}}\epsilon_{\rm{el}}}. \label{eq:3.9AA} 
\end{eqnarray}  
  
Substituting Eqs.~(\ref{eq:3.8AA}) and~(\ref{eq:3.5}) into Eq.~(\ref{eq:3.9AA}) will lead us to Eq.~(\ref{eq:3.11}) as shown below, 

\begin{eqnarray}
\frac{1}{\epsilon} &&= \frac{1}{1 - (\omega(\textbf{k})^2/\omega^2)} \cdot \frac{1}{1 + (\textsl{K}_s^2/k^2)\exp[\lambda(E_F^0 - \xi)]} \nonumber \\&& = \bigg(\frac{1}{1 + (\textsl{K}_s^2/k^2)e^{\lambda(E_F^0 - \xi)}}\bigg)\bigg(\frac{\omega^2}{\omega^2 - \omega(\textbf{k})^2}\bigg). \label{eq:3.10AA} 
\end{eqnarray}  

\subsection{Derivation of Eq.~(\ref{eq:3.15})}\label{C}

Using Eq.~(\ref{eq:3.12}), the equations of motion can be written as

\begin{eqnarray}
&M_1\ddot{u}_1(na)& = -\frac{\partial U_{\rm{harm}}}{\partial u_1(na)} \nonumber \\&& = -\frac{Q}{2}\cdot 2 [u_1(na) - u_2(na)] -\frac{G}{2}\cdot 2 [u_2(na) - u_1((n+1)a)] \nonumber \\&& = -Q[u_1(na) - u_2(na)] -G[u_2(na) - u_1((n+1)a)], \label{eq:3.11AA} 
\end{eqnarray}

\begin{eqnarray}
&M_2\ddot{u}_2(na)& = -\frac{\partial U_{\rm{harm}}}{\partial u_2(na)} \nonumber \\&& = -\frac{Q}{2}\cdot 2 [u_1(na) - u_2(na)](-1) - \frac{G}{2}\cdot 2 [u_2(na) - u_1((n+1)a)] \nonumber \\&& = -Q[u_2(na) - u_1(na)] -G[u_2(na) - u_1((n+1)a)]. \label{eq:3.12AA}
\end{eqnarray}
   
The dressed harmonic potential energy ($\phi^{\rm{dressed}}(x)$) and the interaction potential constants (Q and G) are respectively, given by

\begin{eqnarray}
\phi^{\rm{dressed}}(x) = \phi(x)\exp[\lambda(\xi - E_F^0)], \label{eq:3.13AA}
\end{eqnarray}

\begin{eqnarray}
Q^{\rm{dressed}} = \frac{\partial^2 \phi^{\rm{dressed}}[(na),x]}{\partial x^2} = Q\exp[\lambda(\xi - E_F^0)], \label{eq:3.14AA}
\end{eqnarray}
 
\begin{eqnarray}
G^{\rm{dressed}} = \frac{\partial^2 \phi^{\rm{dressed}}[(na + d),x]}{\partial x^2} = G\exp[\lambda(\xi - E_F^0)]. \label{eq:3.15AA}
\end{eqnarray}

Here, the two ions with their respective masses, $M_1$ and $M_2$ have been considered, in which the ionic oscillations are harmonic ($d \leq a/2$). The solutions we need are in the form of   

\begin{eqnarray}
u_1(na) = b_1e^{i(kna - \omega t)}, ~u_2(na) = b_2 e^{i(kna - \omega t)}, \label{eq:3.16AA}
\end{eqnarray}

where $k$ is the wavevector, while $b_1$ and $b_2$ are constants. By substituting Eq.~(\ref{eq:3.16AA}) into Eq.~(\ref{eq:3.11AA}) and~(\ref{eq:3.12AA}), one can obtain two coupled equations:

\begin{eqnarray}
[M_1\omega^2-(Q+G)e^{\lambda(\xi - E_F^0)}]b_1  + [Q+Ge^{ika}]b_2 e^{\lambda(\xi - E_F^0)} = 0, \label{eq:3.17AA}
\end{eqnarray}

\begin{eqnarray}
[Q+Ge^{ika}]b_2 e^{\lambda(\xi - E_F^0)} + [M_2\omega^2-(Q+G)e^{\lambda(\xi - E_F^0)}]b_1  = 0. \label{eq:3.18AA}
\end{eqnarray}

These homogeneous equations can be solved by setting the determinant equal to zero, i.e.,

\begin{eqnarray}
&&[M_1\omega^2-(Q+G)e^{\lambda(\xi - E_F^0)}][M_2\omega^2-(Q+G)e^{\lambda(\xi - E_F^0)}] = [(Q+Ge^{ika})b_2 e^{\lambda(\xi - E_F^0)}]^2. \nonumber \\&& \label{eq:3.19AA}
\end{eqnarray}

Therefore,  

\begin{eqnarray}
&&[M_1M_2]\omega^4 + [-(Q+G)(M_1+M_2)e^{\lambda(\xi - E_F^0)}]\omega^2 \nonumber \\&& + [(Q+G)^2 - Q^2 - G^2 - 2QG\cos(ka)]e^{2\lambda(\xi - E_F^0)} = 0. \label{eq:3.20AA}
\end{eqnarray}

By identifying $\textbf{A} = M_1M_2$, $\textbf{B} = -(Q+G)(M_1+M_2)e^{\lambda(\xi - E_F^0)}$ and $\textbf{C} = (Q+G)^2 - Q^2 - G^2 - 2QG\cos(ka)]e^{2\lambda(\xi - E_F^0)}$, we can obtain Eq.~(\ref{eq:3.15}) from Eq.~(\ref{eq:3.21AA}) given below: 

\begin{eqnarray}
\omega^2_{\pm} = -\frac{\textbf{B} \pm \sqrt{\textbf{B}^2-4\textbf{AC}}}{2\textbf{A}}. \label{eq:3.21AA}
\end{eqnarray}
  
\subsection{Derivation of Eq.~(\ref{eq:3.23})}\label{D}

From Eq.~(\ref{eq:3.17}), $E_n^{(0)}$ and $E_m^{(0)}$ are the zeroth order initial and excited states, respectively. As such, their difference due to phonon absorption is given by

\begin{eqnarray}
E_m^{(0)} - E_n^{(0)} = E(\textbf{k}^*) + \hbar\omega(\textbf{k} - \textbf{k}^*) - E(\textbf{k}). \label{eq:3.22AA}
\end{eqnarray}

Prior to phonon absorption, one can identify the excited states as unoccupied, while the initial states as occupied. Thus, we need to include a probability factor in the form of $\sum_{\textbf{k},\textbf{k}^*}n_{\textbf{k}}(1 - n_{\textbf{k}^*})$, which satisfies the Bose-Einstein statistics. As a consequence, we can arrive at Eq.~(\ref{eq:3.18}). The next step is to obtain Eq.~(\ref{eq:3.20}):

\begin {eqnarray}
&&V_{\textbf{k},\textbf{k}^*} = \frac{\partial^2E^{(2)}}{\partial n_\textbf{k}\partial n_{\textbf{k}^*}} \nonumber \\&& = \frac{\partial^2}{\partial n_\textbf{k}\partial n_{\textbf{k}^*}} \sum_{\textbf{k},\textbf{k}^*}n_{\textbf{k}}(1 - n_{\textbf{k}^*}) \frac{\left|\left\langle \varphi_{\textbf{k}}\left|H_{\rm{ep}}\right|\varphi_{\textbf{k}^*} \right\rangle\right|^2}{E(\textbf{k}) - E(\textbf{k}^*) - \hbar\omega(\textbf{k} - \textbf{k}^*)} \nonumber \\&& = |g_{\textbf{k},\textbf{k}^*}|^2\frac{\partial^2}{\partial n_\textbf{k}\partial n_{\textbf{k}^*}} \bigg[\frac{n_{\textbf{k}}(1 - n_{\textbf{k}^*})}{E(\textbf{k}) - E(\textbf{k}^*) - \hbar\omega(\textbf{k} - \textbf{k}^*)} + \frac{n_{\textbf{k}^*}(1 - n_{\textbf{k}})}{E(\textbf{k}^*) - E(\textbf{k}) - \hbar\omega(\textbf{k}^* - \textbf{k})}\bigg] \nonumber \\&& = |g_{\textbf{k},\textbf{k}^*}|^2\frac{\partial}{\partial n_{\textbf{k}^*}} \bigg[\frac{(1 - n_{\textbf{k}^*})}{E(\textbf{k}) - E(\textbf{k}^*) - \hbar\omega(\textbf{k} - \textbf{k}^*)} + \frac{n_{\textbf{k}^*}}{E(\textbf{k}^*) - E(\textbf{k}) - \hbar\omega(\textbf{k}^* - \textbf{k})}\bigg] \nonumber \\&& = -|g_{\textbf{k},\textbf{k}^*}|^2 \bigg[\frac{1}{E(\textbf{k}) - E(\textbf{k}^*) - \hbar\omega(\textbf{k} - \textbf{k}^*)} + \frac{1}{E(\textbf{k}^*) - E(\textbf{k}) - \hbar\omega(\textbf{k}^* - \textbf{k})}\bigg] \nonumber \\&& =
|g_{\textbf{k},\textbf{k}^*}|^2\bigg[\frac{2\hbar\omega(\textbf{k}-\textbf{k}^*)e^{\frac{1}{2}\lambda(\xi - E_F^0)}}{[\hbar\omega(\textbf{k}-\textbf{k}^*)e^{\frac{1}{2}\lambda(\xi - E_F^0)}]^2 - [E(\textbf{k}) - E(\textbf{k}^*)]^2}\bigg], \nonumber \\&& \label{eq:3.23AA}
\end {eqnarray}  

where $|g_{\textbf{k},\textbf{k}^*}|^2 = |g_{\textbf{k}^*,\textbf{k}}|^2$. After dressing Eq.~(\ref{eq:3.21}) with ionization energy, one obtains

\begin {eqnarray}
\left|g_{\textbf{k},\textbf{k}^*}\right|^2 = \frac{1}{V}\frac{e^2}{\epsilon_0\big[|\textbf{k}-\textbf{k}^*|^2 + K_s^2e^{\lambda(\xi - E_F^0)}\big]}\frac{1}{2}\hbar\omega(\textbf{k}-\textbf{k}^*)e^{\frac{1}{2}\lambda(\xi - E_F^0)}. \label{eq:3.24AA}
\end {eqnarray}
  
Substituting Eq.~(\ref{eq:3.24AA}) into Eq.~(\ref{eq:3.23AA}) leads to

\begin {eqnarray}
&&V_{\rm{ep}}(\textbf{k},\textbf{k}^*) = \frac{e^2}{2V\epsilon_0}\frac{\hbar\omega(\textbf{k}-\textbf{k}^*)e^{\frac{1}{2}\lambda(\xi - E_F^0)}}{{\big[|\textbf{k}-\textbf{k}^*|^2 + K_s^2e^{\lambda(\xi - E_F^0)}\big]}}\bigg[\frac{2\hbar\omega(\textbf{k}-\textbf{k}^*)e^{\frac{1}{2}\lambda(\xi - E_F^0)}}{[\hbar\omega(\textbf{k}-\textbf{k}^*)e^{\frac{1}{2}\lambda(\xi - E_F^0)}]^2 - [E(\textbf{k}) - E(\textbf{k}^*)]^2}\bigg]. \nonumber \\&& \label{eq:3.25AA}
\end {eqnarray} 

Now, invoking the effective mass theorem [Eq.~(\ref{eq:3.22})], Eq.~(\ref{eq:3.25AA}) can be written as given below to obtain Eq.~(\ref{eq:3.23}). 

\begin {eqnarray}
&&V_{\rm{ep}}(\textbf{k},\textbf{k}^*) = \frac{e^2}{2V\epsilon_0}\frac{\hbar\omega(\textbf{k}-\textbf{k}^*)e^{\frac{1}{2}\lambda(\xi - E_F^0)}}{{\big[|\textbf{k}-\textbf{k}^*|^2 + K_s^2e^{\lambda(\xi - E_F^0)}\big]}}\bigg[\frac{2\hbar\omega(\textbf{k}-\textbf{k}^*)e^{\frac{1}{2}\lambda(\xi - E_F^0)}}{[\hbar\omega(\textbf{k}-\textbf{k}^*)e^{\frac{1}{2}\lambda(\xi - E_F^0)}]^2}\bigg] \nonumber \\&& = \frac{1}{V\epsilon_0}\bigg[\frac{e^2}{|\textbf{k}-\textbf{k}^*|^2 + K_s^2\exp[\lambda(E_F^0 - \xi)]}\bigg]. \label{eq:3.26AA}
\end {eqnarray}  

\subsection{Derivation of Eq.~(\ref{eq:3.26})}\label{E}

Equation~(\ref{eq:3.15b}) can also be written as

\begin {eqnarray}
&U& = U_{\rm{stat}} + \frac{1}{V}\sum_{\textbf{k},s}\frac{1}{2}\hbar\omega_s(\textbf{k})e^{\frac{1}{2}\lambda(\xi - E_F^0)} + \frac{1}{V}\sum_{\textbf{k},s}\hbar\omega_s(\textbf{k})e^{\frac{1}{2}\lambda(\xi - E_F^0)}n_\textbf{k}. \label{eq:3.27AA}
\end {eqnarray} 

After substituting Eq.~(\ref{eq:3.19}) into Eq.~(\ref{eq:3.27AA}), one can arrive at Eq.~(\ref{eq:3.24}). Whereas, the $C_v$ can be derived from Eq.~(\ref{eq:3.27AA}) as

\begin {eqnarray}
&&C_v = \frac{\partial U}{\partial T} = \frac{1}{V}\sum_{\textbf{k},s}\frac{\partial}{\partial T}\frac{\hbar\omega_s(\textbf{k})e^{\frac{1}{2}\lambda(\xi - E_F^0)}}{e^{\beta\hbar\omega_s(\textbf{k})e^{\frac{1}{2}\lambda(\xi - E_F^0)}} - 1} \nonumber \\&& = \sum_{s}\frac{\partial}{\partial T}\int\frac{d\textbf{k}}{(2\pi)^3}\frac{\hbar\omega_s(\textbf{k})e^{\frac{1}{2}\lambda(\xi - E_F^0)}}{e^{\beta\hbar\omega_s(\textbf{k})e^{\frac{1}{2}\lambda(\xi - E_F^0)}} - 1}. \label{eq:3.28AA}
\end {eqnarray}

Next, we substitute these equations given below into Eq.~(\ref{eq:3.28AA}):  

\begin {eqnarray}
\omega(\textbf{k}) = c_s(\hat{\textbf{k}})ke^{\frac{1}{2}\lambda(\xi - E_F^0)}, ~d\textbf{k} = k^2dk~ \sin\theta ~d\theta ~d\phi = k^2dkd\Omega,\label{eq:3.29AA} 
\end {eqnarray}

\begin {eqnarray}
x = \beta\hbar c_ske^{\frac{1}{2}\lambda(\xi - E_F^0)},~ \frac{dx}{dk} = \beta\hbar c_s e^{\frac{1}{2}\lambda(\xi - E_F^0)}. \label{eq:3.30AA}
\end {eqnarray}

Hence, we get

\begin {eqnarray}
&&C_v = \sum_{s}\frac{\partial}{\partial T}\int\frac{k^2dk~d\Omega}{(2\pi)^3}\frac{\hbar c_ske^{\frac{1}{2}\lambda(\xi - E_F^0)}}{e^{\beta\hbar c_ske^{\frac{1}{2}\lambda(\xi - E_F^0)}} - 1} \nonumber \\&& = \sum_{s}\frac{\partial}{\partial T}\int\frac{d\Omega}{4\pi}\int\frac{k^2dk}{2\pi^2}\frac{\hbar c_ske^{\frac{1}{2}\lambda(\xi - E_F^0)}}{e^{\beta\hbar c_ske^{\frac{1}{2}\lambda(\xi - E_F^0)}} - 1} \nonumber \\&& = \frac{1}{2\pi^2}\sum_{s}\frac{\partial}{\partial T}\int\frac{d\Omega}{4\pi}\int\bigg[\frac{x^2dx}{\big[\beta \hbar c_se^{\frac{1}{2}\lambda(\xi - E_F^0)}\big]^3}\bigg]\bigg[\frac{x/\beta}{e^x - 1}\bigg] \nonumber \\&& = \bigg[\frac{3}{2\pi^2}\frac{\partial}{\partial T}\bigg]\cdot \bigg[\frac{1}{3}\sum_s \int\frac{d\Omega}{4\pi}\cdot \frac{1}{c_s}\bigg]\cdot \bigg[\frac{1}{\beta^4 \hbar^3 e^{\frac{3}{2}\lambda(\xi - E_F^0)}} \bigg]\int_0^{\infty} \frac{x^3dx}{e^x - 1}. \label{eq:3.31AA}
\end {eqnarray}

Using

\begin {eqnarray}
\frac{1}{c^3} = \frac{1}{3}\sum_s\int\frac{d\Omega}{4\pi}\cdot \frac{1}{c_s^3}, \label{eq:3.32AA}
\end {eqnarray}

\begin {eqnarray}
\int_0^{\infty}\frac{x^3dx}{e^x - 1} = \sum_{n=1}^{\infty}\int_0^{\infty}x^3e^{-nx}dx = 6\sum_{n=1}^{\infty}\frac{1}{n^4} = \frac{\pi^4}{15}, \label{eq:3.33AA}
\end {eqnarray}

we obtain Eq.~(\ref{eq:3.26}),

\begin {eqnarray}
&&C_v = \frac{\pi^2k_B}{10}\frac{\partial T^4}{\partial T}\bigg[\frac{k_B}{\hbar c}\bigg]^3e^{-\frac{3}{2}\lambda(\xi - E_F^0)} \nonumber \\&& = \frac{2\pi^2k_B}{5}\bigg[\frac{k_BT}{\hbar c}\bigg]^3e^{-\frac{3}{2}\lambda(\xi - E_F^0)}. \label{eq:3.34AA}
\end {eqnarray}

Alternatively, instead of using Eq.~(\ref{eq:3.30AA}), we use

\begin {eqnarray}
x = \beta\hbar c_sk,~ \frac{dx}{dk} = \beta\hbar c_s, \label{eq:3.35AA}
\end {eqnarray}

to get [also after using Eq.~(\ref{eq:3.32AA})]

\begin {eqnarray}
&&C_v = \frac{\partial}{\partial T}\sum_s\int\frac{d\Omega}{4\pi}\int\frac{k^2 dk}{2\pi^2}\frac{\hbar c_s k e^{\frac{1}{2}\lambda(\xi - E_F^0)}}{e^{\beta \hbar c_s k \exp\big[\frac{1}{2}\lambda(\xi - E_F^0)\big]} - 1} \nonumber \\&& = \frac{1}{2\pi^2}\frac{\partial}{\partial T}\sum_s\int\frac{d\Omega}{4\pi}\int\bigg[\frac{x}{\beta \hbar c_s}\bigg]^2\cdot \bigg[\frac{dx}{\beta \hbar c_s}\bigg] \frac{xe^{\frac{1}{2}\lambda(\xi - E_F^0)}}{\beta [e^{x\exp\big[\frac{1}{2}\lambda(\xi - E_F^0)\big]} - 1]} \nonumber \\&& = \frac{3}{2\pi^2}\frac{e^{\frac{1}{2}\lambda(\xi - E_F^0)}}{\hbar^3 c_s^3}\frac{\partial (1/\beta^4)}{\partial T}\int_0^{\infty}\frac{x^3dx}{e^{x\exp\big[\frac{1}{2}\lambda(\xi - E_F^0)\big]} - 1}. \label{eq:3.36AA}
\end {eqnarray}

Substituting 

\begin {eqnarray}
&&\int_0^{\infty}\frac{x^3dx}{e^{x\exp\big[\frac{1}{2}\lambda(\xi - E_F^0)\big]} - 1} = \sum_{n=1}^{\infty}\int_0^{\infty}x^3\exp\big[-nxe^{\frac{1}{2}\lambda(\xi - E_F^0)}\big]dx = 3!\bigg[\frac{1}{e^{\frac{1}{2}\lambda(\xi - E_F^0)}}\bigg]^4\sum_{n=1}^{\infty}\frac{1}{n^4} \nonumber \\&& = \frac{\pi^4}{15}\frac{1}{e^{2\lambda(\xi - E_F^0)}}, \nonumber \\&& \label{eq:3.37AA}
\end {eqnarray}

into Eq.~(\ref{eq:3.36AA}), we can see that Eq.~(\ref{eq:3.36AA}) becomes Eq.~(\ref{eq:3.34AA}), which is Eq.~(\ref{eq:3.26}).  

\subsection{Derivation of Eq.~(\ref{eq:3.30})}\label{F}

From the left-hand side (LHS) equation of Eq.~(\ref{eq:3.29AA}),   

\begin {eqnarray}
k = \frac{\omega}{c}e^{-\frac{1}{2}\lambda(\xi - E_F^0)}. \label{eq:3.38AA} 
\end {eqnarray}

On the other hand, the number of phonon modes, $N$ is given by

\begin {eqnarray}
N = \frac{(4\pi/3)k^3}{(2\pi/L)^3} = \frac{Vk^3}{6\pi^2}, \label{eq:3.39AA} 
\end {eqnarray}

where $L$ and $V$ are the length and volume, respectively. Next, by substituting Eq.~(\ref{eq:3.38AA}) into Eq.~(\ref{eq:3.39AA}), 

\begin {eqnarray}
N = \frac{V\omega^3}{6\pi^2c^3}e^{-\frac{3}{2}\lambda(\xi - E_F^0)}, \label{eq:3.40AA} 
\end {eqnarray}

and differentiating Eq.~(\ref{eq:3.40AA}) with respect to $\omega$ will lead us directly to Eq.~(\ref{eq:3.28}). Substituting Eq.~(\ref{eq:3.38AA}) into Eq.~(\ref{eq:3.40AA}) gives

\begin {eqnarray}
&&\frac{N}{V}6\pi^2 = \frac{\omega^3}{c^3}e^{-\frac{3}{2}\lambda(\xi - E_F^0)} = k^3e^{-\frac{3}{2}\lambda(\xi - E_F^0)}, \nonumber \\&& 
\frac{N}{V}6\pi^2e^{\frac{3}{2}\lambda(\xi - E_F^0)} = k^3. \label{eq:3.41AA}
\end {eqnarray}
 
Subsequently, we define, $k = k_D$ and $\omega = \omega_D$, thus 

\begin {eqnarray}
k_B\Theta_D = \hbar\omega_D = \hbar c k_D, ~ \Theta_D = \frac{\hbar c}{k_B}k_D. \label{eq:3.42AA} 
\end {eqnarray}

Inserting Eq.~(\ref{eq:3.41AA}) into Eq.~(\ref{eq:3.42AA}) gives 

\begin {eqnarray}
\Theta_D = \frac{\hbar c}{k_B}\bigg[\frac{6\pi^2Ne^{\frac{3}{2}\lambda(\xi - E_F^0)}}{V}\bigg]^{\frac{1}{3}}, \label{eq:3.43AA} 
\end {eqnarray}

which is equivalent to Eq.~(\ref{eq:3.30}) by noting that $n_{\rm{ph}} = N/V$. 

\subsection{Derivation of Eq.~(\ref{eq:3.32})}\label{G}

From Eq.~(\ref{eq:3.250}), noting $\lim_{V \rightarrow \infty}(1/V)\sum_\textbf{k}$ = $\int d\textbf{k}/(2\pi)^3$, and using the RHS equation of Eq.~(\ref{eq:3.29AA}) one gets 

\begin {eqnarray}
&C_v& = \frac{3\hbar c}{2\pi^2}\int^{k_{\rm{D}}}_0 k^3dke^{\frac{1}{2}\lambda(\xi - E_F^0)}\frac{\partial}{\partial T}\bigg[\frac{1}{e^{\beta\hbar cke^{\frac{1}{2}\lambda(\xi - E_F^0)}} - 1}\bigg], \label{eq:3.44AA}
\end {eqnarray}

where

\begin {eqnarray}
c = \bigg[\frac{1}{3}\sum_s \int \frac{d\Omega}{4\pi}c_s(\hat{\textbf{k}})\bigg]^{\rm{isotropic}}_{\rm{polarization}}, \label{eq:3.45AA}
\end {eqnarray}

\begin {eqnarray}
&&\frac{\partial}{\partial T}\bigg[\frac{1}{e^{\beta\hbar cke^{\frac{1}{2}\lambda(\xi - E_F^0)}} - 1}\bigg] = \frac{\partial}{\partial T}\big[e^{\beta\hbar cke^{\frac{1}{2}\lambda(\xi - E_F^0)}} - 1\big]^{-1} \nonumber \\&& = -1\big[e^{\beta\hbar cke^{\frac{1}{2}\lambda(\xi - E_F^0)}} - 1\big]^{-2}\bigg[-\frac{\hbar c k}{k_BT^2}e^{\beta\hbar cke^{\frac{1}{2}\lambda(\xi - E_F^0)}}\bigg] \nonumber \\&& = \frac{\hbar c k}{k_BT^2}\frac{e^{\beta\hbar cke^{\frac{1}{2}\lambda(\xi - E_F^0)}}}{\big[e^{\beta\hbar cke^{\frac{1}{2}\lambda(\xi - E_F^0)}} - 1 \big]^2}. \label{eq:3.46AA}
\end {eqnarray}

Therefore, we can derive $C_v$ as given in Eq.~(\ref{eq:3.31}). Consequently, after the change of variables as given in Eqs.~(\ref{eq:3.29AA}) and~(\ref{eq:3.30AA}),        

\begin {eqnarray}
&C_v& = \frac{3(\hbar c)^2}{2\pi^2k_BT^2}\int^{k_{\rm{D}}}_0 k^4dke^{\lambda(\xi - E_F^0)} \frac{e^{\beta\hbar cke^{\frac{1}{2}\lambda(\xi - E_F^0)}}}{[e^{\beta\hbar cke^{\frac{1}{2}\lambda(\xi - E_F^0)}} - 1]^2} \nonumber \\&& = \frac{3(\hbar c)^2}{2\pi^2k_BT^2}\int^{x_{\rm{D}}}_0 \frac{(xk_BT)^4e^{\lambda(\xi - E_F^0)}}{(\hbar c)^4e^{2\lambda(\xi - E_F^0)}}\frac{dxk_BT}{\hbar c e^{\frac{1}{2}\lambda(\xi - E_F^0)}}\frac{e^{\beta\hbar cke^{\frac{1}{2}\lambda(\xi - E_F^0)}}}{[e^{\beta\hbar cke^{\frac{1}{2}\lambda(\xi - E_F^0)}} - 1]^2} \nonumber \\&& = \frac{3(\hbar c)^2}{2\pi^2k_BT^2}\bigg[\frac{k_BT}{\hbar c}\bigg]^5 e^{(1-2-\frac{1}{2})\lambda(\xi - E_F^0)}\int^{x_{\rm{D}}}_0 \frac{x^4e^x dx}{\big(e^x -1\big)^2}.  \label{eq:3.47AA}
\end {eqnarray}

From Eq.~(\ref{eq:3.42AA}), one gets $\hbar c = (k_B\Theta_D/k_D)$. Substituting this and Eq.~(\ref{eq:3.39AA}) into Eq.~(\ref{eq:3.47AA}), and after some algebraic rearrangements, one can arrive at

\begin {eqnarray}
&C_v& = \frac{3(\hbar c)^2}{2\pi^2k_BT^2}\bigg[\frac{k_BT}{\hbar c}\bigg]^5 e^{(1-2-\frac{1}{2})\lambda(\xi - E_F^0)}\int^{x_{\rm{D}}}_0 \frac{x^4e^x dx}{\big(e^x -1\big)^2} \nonumber \\&& = \frac{3k_B^4T^3}{2\pi^2(\hbar c)^3} e^{-\frac{3}{2}\lambda(\xi - E_F^0)}\int^{x_{\rm{D}}}_0 \frac{x^4e^x dx}{\big(e^x -1\big)^2} \nonumber \\&& = \frac{3k_B}{2\pi^2}\big(6\pi^2n_{\rm{ph}}\big) \bigg[\frac{T}{\Theta_De^{\frac{1}{2}\lambda(\xi - E_F^0)}}\bigg]^3\int^{x_{\rm{D}}}_0 \frac{x^4e^x dx}{\big(e^x -1\big)^2}, \label{eq:3.48AA}
\end {eqnarray}
 
which is nothing but Eq.~(\ref{eq:3.32}). Again, using Eq.~(\ref{eq:3.33AA}) and $x_D = \beta\hbar c k_D = (k_B\Theta_D/k_BT) = \Theta_D/T \rightarrow \infty$ for $T < \Theta_D$, one can solve Eq.~(\ref{eq:3.48AA}):

\begin {eqnarray}
\int_0^{\infty}\frac{x^4e^xdx}{(e^x - 1)^2} = \sum_{n=1}^{\infty}\int_0^{\infty}x^4e^{n(x-2x)}dx = 4!\sum_{n=1}^{\infty}\frac{1}{n^4} = \frac{24\pi^4}{90}. \label{eq:3.49AA}
\end {eqnarray}

Therefore,

\begin {eqnarray}
&C_v& = \frac{12\pi^4}{5}n_{\rm{ph}}k_B \bigg[\frac{T}{\Theta_De^{\frac{1}{2}\lambda(\xi - E_F^0)}}\bigg]^3. \label{eq:3.50AA}
\end {eqnarray}

If we use Eq.~(\ref{eq:3.35AA}) instead, then we need to invoke Eq.~(\ref{eq:3.37AA}) and the result [Eq.~(\ref{eq:3.50AA})] remains the same: from Eq.~(\ref{eq:3.47AA})

\begin {eqnarray}
&C_v& = \frac{3(\hbar c)^2}{2\pi^2k_BT^2}\int^{k_{\rm{D}}}_0 k^4dke^{\lambda(\xi - E_F^0)} \frac{e^{\beta\hbar cke^{\frac{1}{2}\lambda(\xi - E_F^0)}}}{[e^{\beta\hbar cke^{\frac{1}{2}\lambda(\xi - E_F^0)}} - 1]^2} \nonumber \\&& = \frac{3(\hbar c)^2}{2\pi^2k_BT^2}e^{\lambda(\xi - E_F^0)}\int^{x_{\rm{D}}}_0 \bigg[\frac{x}{\beta\hbar c}\bigg]^4 \cdot \bigg[\frac{dx}{\beta\hbar c}\bigg]\frac{e^{xe^{\frac{1}{2}\lambda(\xi - E_F^0)}}}{[e^{xe^{\frac{1}{2}\lambda(\xi - E_F^0)}} - 1]^2} \nonumber \\&& = \frac{3k_B^4T^3}{2\pi^2(\hbar c)^3}e^{\lambda(\xi - E_F^0)}\int^{x_{\rm{D}}}_0 \frac{x^4e^{xe^{\frac{1}{2}\lambda(\xi - E_F^0)}}}{[e^{xe^{\frac{1}{2}\lambda(\xi - E_F^0)}} - 1]^2}dx. \label{eq:3.51AA}
\end {eqnarray}

Using $\hbar c = (k_B\Theta_D/k_D)$ and Eq.~(\ref{eq:3.39AA}),

\begin {eqnarray}
&C_v& = \frac{3}{2\pi^2}n_{\rm{ph}}6\pi^2k_Be^{\lambda(\xi - E_F^0)}\frac{T^3}{\Theta_D^3}\int^{\infty}_0 \frac{x^4e^{xe^{\frac{1}{2}\lambda(\xi - E_F^0)}}}{[e^{xe^{\frac{1}{2}\lambda(\xi - E_F^0)}} - 1]^2}dx. \label{eq:3.52AA}
\end {eqnarray}

From,

\begin {eqnarray}
&&\int_0^{\infty}\frac{x^4dx}{\big(e^{x\exp\big[\frac{1}{2}\lambda(\xi - E_F^0)\big]} - 1\big)^2} = \sum_{n=1}^{\infty}\int_0^{\infty}x^4\exp\big[n(x-2x)e^{\frac{1}{2}\lambda(\xi - E_F^0)}\big]dx = 4!\bigg[\frac{1}{e^{\frac{1}{2}\lambda(\xi - E_F^0)}}\bigg]^5\sum_{n=1}^{\infty}\frac{1}{n^4} \nonumber \\&& = \frac{24\pi^4}{90}\frac{1}{e^{\frac{5}{2}\lambda(\xi - E_F^0)}}, \nonumber \\&& \label{eq:3.53AA}
\end {eqnarray}

we can rewrite Eq.~(\ref{eq:3.52AA}) as

\begin {eqnarray}
&C_v& = \frac{12\pi^4}{5}n_{\rm{ph}}k_B e^{(1-\frac{5}{2})\lambda(\xi - E_F^0)}\bigg[\frac{T}{\Theta_D}\bigg]^3, \label{eq:3.54AA}
\end {eqnarray}

which is Eq.~(\ref{eq:3.50AA}).

\end{document}